\documentclass[journal, onecolumn, 10pt]{IEEEtran}
\usepackage{epsfig,rotating,setspace,latexsym,amsmath,epsf,amssymb,bm}
\usepackage{cite,graphicx,color,hyperref}
\usepackage{booktabs}
\usepackage{amsmath,graphicx}
\usepackage{rotating,latexsym,amsmath,amssymb,bm}
\usepackage{cite}
\usepackage{algpseudocode}
\usepackage{algorithm}
\usepackage{algorithmicx}
\usepackage{xfrac}
\usepackage{amsthm}
\usepackage{subfigure}
\usepackage{graphicx}
\usepackage{subfigure}
\usepackage{algpseudocode}
\usepackage{algorithm}
\usepackage{balance}
\usepackage{stfloats}

\newtheorem{thm}{Theorem}
\newtheorem{corr}[thm]{Corollary}
\theoremstyle{remark}
\newtheorem{exmp}{Example}
\theoremstyle{definition}
\newtheorem{defn}{Definition}
\theoremstyle{remark}
\newtheorem{remark}{Remark}
\newcommand{\myeq}[1]{\mathrel{\overset{\makebox[0pt]{\mbox{\normalfont\tiny\sffamily #1}}}{=}}}
\newcommand{\myleq}[1]{\mathrel{\overset{\makebox[0pt]{\mbox{\normalfont\tiny\sffamily #1}}}{\leq}}}

\newcommand{\N}{$N$}
\newcommand{\K}{\mathcal{K}}

\newcommand\blfootnote[1]{%
  \begingroup
  \renewcommand\thefootnote{}\footnote{#1}%
  \addtocounter{footnote}{-1}%
  \endgroup
}

\begin{document}
\sloppy
\title{Fundamental Limits of Caching with Secure Delivery} 
\author{Avik Sengupta, Ravi Tandon, and T.~Charles Clancy\\
\thanks{A. Sengupta and T. C. Clancy are with the Hume Center for National Security and Technology, Department of Electrical and Computer Engineering, Virginia Tech, Blacksburg, VA USA. Email: \{aviksg, tcc\}@vt.edu. R. Tandon is with Discovery Analytics Center, Department of Computer Science, Virginia Tech, Blacksburg, VA USA.	Email: tandonr@vt.edu.}\vspace{-5pt}}
\maketitle 
\thispagestyle{plain}
\pagestyle{plain}

\begin{abstract}
Caching is emerging as a vital tool for alleviating the severe capacity crunch in modern content-centric wireless networks. The main idea behind caching is to store parts of popular content in end-users' memory and leverage the locally stored content to reduce peak data rates. By jointly designing content placement and delivery mechanisms, recent works have shown order-wise reduction in transmission rates in contrast to traditional methods. 
In this work, we consider the \textit{secure caching problem} with the additional goal of minimizing information leakage to an external wiretapper. The fundamental cache memory vs. transmission rate trade-off for the secure caching problem is characterized. Rather surprisingly, these results show that security can be introduced at a negligible cost, particularly for large number of files and users. It is also shown that the rate achieved by the proposed caching scheme with secure delivery is within a constant multiplicative factor from the information-theoretic optimal rate for almost all parameter values of practical interest.
\end{abstract}

\vspace{-18pt}
\section{Introduction}
\label{sec:intro}
\blfootnote{Parts of this work was presented at IEEE ICC Wireless Physical Layer Security Workshop, June 2014\nocite{aviksg-icc} and at IEEE International Symposium on Information Theory, July 2014\nocite{aviksg-isit}.}
In modern content-centric wireless networks, caching helps in reducing the peak network load at times of high traffic volume. Fractions of popular content are stored locally in end-users' cache memories distributed across a given network. At times of high demand, the users can be partly served locally from their cache, thereby reducing the network load. Caching generally works in two phases - the \textit{storage phase} and the \textit{delivery phase}. The general caching problem has been well studied in literature \cite{Dowdy,almeroth,Korupolu,Baev}. Traditionally, the delivery phase of caching systems operate as a series of dedicated unicast transmissions to individual users by transmitting fractions of requested files which are not stored in their caches. However, this is not a scalable solution as the number of users in the system increases. A more efficient solution is to deliver content simultaneously to users through multicast transmissions. Most of the prior works in this area tend to use a fixed delivery scheme and then optimize the storage phase to suit the delivery scheme \cite{Korupolu,Baev}. Further, their investigations are mainly based on the gains obtained from local content distribution, ignoring the global cache interactions and content sharing as a factor for extracting caching gain. 

More recently, \cite{Maddah-Ali,Maddah-Ali-decentralized,Maddah-Ali-nonuniform,Maddah-Ali-Online,Maddah-Ali-Delay-sensitive,diggavi} have proposed information theoretic formulations of the caching problem. In \cite{Maddah-Ali}, a scheme is proposed which, in addition to the local caching gain, is also capable of offering a global caching gain. The scheme takes the cumulative size of the network cache memory into consideration and \textit{jointly designs} the cache storage phase and a coded mutlicast delivery phase. This achieves a global caching gain which provides an order-wise improvement over local caching gain. The fundamental concepts presented in \cite{Maddah-Ali} are extended to the case of decentralized storage in \cite{Maddah-Ali-decentralized} and non-uniform ZipF \cite{zipf_dist} user demands in \cite{Maddah-Ali-nonuniform,ISWCS_Ji}. Some extensions of the caching problem have been investigated in the case of Device-to-Device (D2D) communications in \cite{Molisch-onecache,Molisch_D2D_distcache,FundD2D,fund_ji}, from the perspective of content distribution networks in \cite{Femto-journal} and reinforcement learning in \cite{Gunduz_ML,Gunduz_ISIT,aviksg-iswcs}. 

In this paper, we investigate the fundamental\textit{ security} aspects of the caching problem in the presence of an external adversary (wiretapper). To this end, we introduce the \textit{secure caching problem} in which the multicast communication between the central server and the users (delivery phase) occurs over a \textit{public (insecure) channel}. The defining feature of this problem is to capture the tradeoff between the multicast rate of the insecure link and the size of the cache memory. To the best of our knowledge, none of the works on cache storage and placement design deal with security issues. 
We consider a system with a central server connected to $K$ users through an error-free rate-limited link. The server has a database of $N$ files denoted by $(W_{1}, \ldots , W_{N})$, where each file is of size $F$ bits. For the scope of this paper, we assume that a user can request access to \textit{any} one of the files at a given time. Each user has a cache memory $Z_{k}$ of size $M F$ bits for any real number $M \in [0,N]$. Similar to \cite{Maddah-Ali}, the system operates over two phases: a cache \textit{storage phase} and a \textit{delivery phase}. The storage phase can be of two types: \textit{centralized storage} or \textit{decentralized storage}. In case of centralized storage, the central server stores the cache $Z_{k}$ of user $k$ with some content, which is a function of the files $(W_{1}, \ldots, W_{N})$. In case of decentralized storage, the user $k$ is allowed to store any random combination of bits from each file without coordination from the central server. User $k$ (for $k$$=$$1,\ldots, K$) then requests access to one of the files $W_{d_{k}}$ in the database. In the delivery phase, the central server proceeds by transmitting a signal $X_{(d_{1},\ldots, d_{K})}$ of size $RF$ bits over the shared link. Using the content $Z_{k}$ (of its cache) and the received signal $X_{(d_{1},\ldots, d_{K})}$, the $k-$th user intends to reconstruct the requested file $W_{d_{k}}$. A memory-rate pair $(M,R)$ is \textit{achievable} if for a (per-user) cache size of $MF$ bits, and using rate $RF$ bits, it is possible for each user to decode its requested file for \textit{any} set of requests $(d_1,\ldots, d_K)$. Let $R^{*}(M)$ denote the smallest rate $R$ such that the pair $(M,R)$ is achievable. The function $R^{*}(M)$ is the fundamental \textit{memory-rate} tradeoff for the caching problem. An approximate characterization for $R^{*}(M)$ was provided in \cite{Maddah-Ali,Maddah-Ali-nonuniform,Maddah-Ali-decentralized}.

\begin{figure}[t]
\centering
\includegraphics[width=3.2 in,height=2.3 in]{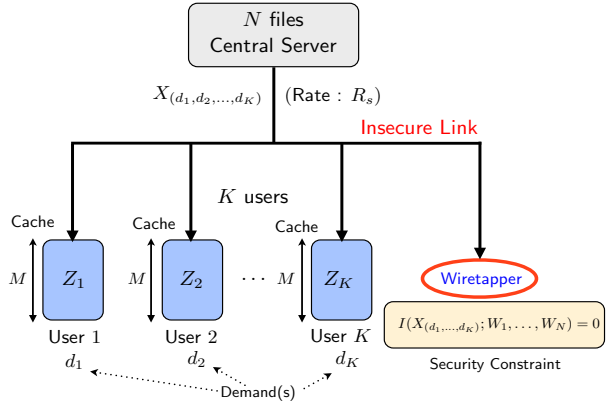}
\vspace{-8pt}
\caption{System Model for Secure Caching.}
\label{fig:sysmod} 
\vspace{-20pt}
\end{figure}

We consider this problem in the presence of an external wiretapper which can observe the multicast communication $X_{(d_{1},\ldots, d_{K})}$ i.e., the communication from the central server to the users occurs over an \textit{insecure} link. {The wiretapper is considered to be strictly out-of-network and is thus able to observe only the multicast delivery which happens over a broadcast channel.} Thus, besides satisfying the users' demands, we require that $X_{(d_1,\ldots,d_K)}$ must not reveal any information about $(W_{{1}}, \ldots, W_{{N}})$ i.e., $I\left(X_{(d_1,\ldots,d_K)};W_1,\ldots,W_N\right) = 0$. As is shown, the additional security constraint necessitates introducing randomness in the form of keys, which occupy a part of the cache of each user. Subsequently, these keys are used in the delivery phase to the keep the delivery information theoretically secure using a one-time-pad scheme \cite{shannon}. {In our system model, the placement phase occurs over unicast channels to individual users and can be secured with the help of individual keys e.g., secure unicast communications using a system similar to code-division-multiple-access (CDMA). As a result, security is considered to be inherent in the placement phase.} Thus, in this work, we consider the security of only the \textit{delivery phase} and not the cache \textit{placement phase}. For this problem, a memory-rate pair $(M,R_{s})$ is \textit{securely achievable} if, for a cache size of $MF$ and a transmission of rate $R_{s}F$ bits, it is possible for each user to decode its requested file and the communication over the shared link reveals no information about any file. Fig. \ref{fig:sysmod} shows the caching system in the presence of a wiretapper. Let $R^{*}_{s}(M)$ denote the smallest $R_{s}$ such that $(M, R_{s})$ is achievable. Thus, the function $R^{*}_{s}(M)$ is the fundamental memory-rate tradeoff for the \textit{secure} caching problem. We investigate both the centralized cache placement as well as the decentralized placement with secure file delivery without any assumptions on user demands and file popularity. 

The main contribution of this paper is an approximate characterization of $R^{*}_{s}(M)$. We design centralized and decentralized caching algorithms which make use of coded multicast delivery to extract global caching gain. The system has uniformly distributed orthogonal keys which are stored across users for secure multicast delivery. We present novel upper and lower bounds on $R^{*}_{s}(M)$ and show that these bounds are within a constant multiplicative gap. Indeed, for a fixed $M$, it is intuitively clear that $R^{*}_{s}(M)\geq R^{*}(M)$, i.e., the minimum rate in presence of a wiretapper must be, in general, larger than in the absence of a wiretapper.  From our results, we show, rather surprisingly, that the cost for incorporating security in both the centralized and decentralized caching schemes is negligible when the number of users and files are large. 
\section{System Model} \label{sec:sysmodel}
Let $(W_{1}, W_{2}, \ldots, W_{N})$ be $N$ independent random variables each uniformly distributed over 
\begin{align}
[2^{F}]\triangleq \{1,2,\ldots, 2^{F}\}
\end{align}
for some $F\in \mathbb{N}$. Each $W_{n}$ represents a file of size $F$ bits. A $(M,R_{s})$ secure caching scheme comprises of $K$
\textit{random} caching functions, $N^{K}$ \textit{random} encoding functions and $KN^{K}$ decoding functions. The $K$ \textit{random} caching functions map the files $(W_{1}, \ldots, W_{N})$ into the cache content:
\begin{align}\vspace{-5pt}
Z_{k}\triangleq \phi_{k}\big(W_{1}, \ldots, W_{N}\big)
\end{align}
for each user $k\in [K]$ during the storage (or placement) phase. The maximum allowable size of the contents of each cache $Z_k$ is $MF$ bits. The $N^{K}$ \textit{random} encoding functions map the 
files $(W_{1}, \ldots, W_{N})$ to the input 
\begin{align}
X_{(d_{1}, \ldots, d_{K})}\triangleq \psi_{(d_{1}, \ldots, d_{K})}\big(W_{1}, \ldots, W_{N}\big)
\end{align}
of the shared link in response to the requests $(d_{1}, \ldots, d_{K})\in [N]^{K}$ during the delivery phase. Finally, the $KN^{K}$ decoding functions map the received signal over the \textit{insecure} shared link $X_{(d_{1}, \ldots, d_{K})}$ and the  cache content $Z_{k}$ to the estimate
\begin{align}
\hat{W}_{(d_{1},\ldots, d_{K}), k}\triangleq \mu_{(d_{1},\ldots, d_{K}), k} \Big(X_{(d_{1}, \ldots, d_{K})}, Z_{k}\Big)
\end{align}
of the requested file $W_{d_{k}}$ for user $k\in [K]$. The probability of error is defined as:
\begin{align}
P_{e}\triangleq \max_{(d_{1}, \ldots, d_{K})\in [N]^{K}} \max_{k\in [K]} \mathbb{P}(\hat{W}_{(d_{1},\ldots, d_{K}), k}\neq W_{d_{k}}).
\end{align}
The information leaked at the wiretapper is defined as:
\begin{align}
L&\triangleq \max_{(d_{1}, \ldots, d_{K})\in [N]^{K}}  I\left(X_{(d_{1}, \ldots, d_{K})}; W_{1}, \ldots, W_{N}\right).
\end{align}\vspace{-10pt}
\begin{defn}\label{def1}
The pair $(M, R_{s})$ is \textit{securely achievable} if for any $\epsilon>0$ and every large enough file size $F$, there exists a $(M, R_{s})$ secure caching scheme with $P_{e}\leq \epsilon$ and $L\leq \epsilon$. We define the secure memory-rate tradeoff 
\begin{align}
R^{*}_{s}(M)\triangleq \inf \{R_{s}: (M, R_{s})\textit{ is securely achievable} \}.
\end{align}
\end{defn}
\section{Centralized Caching with Secure Delivery}
\label{sec:mainres}
The first result gives an achievable rate which upper bounds the optimal memory-rate trade-off $R_s^{*}(M)$ for the centralized caching scheme with secure delivery. Security is incorporated by introducing randomness in the storage and delivery phase of the achievable scheme in form of a set of uniformly distributed orthogonal keys (independent of the data) stored in the cache of each user. The total cache memory (of size $MF$ bits) is divided into two parts - data memory (of size $M_D F$ bits) and key memory (of size $M_K F$ bits) such that $M = M_D + M_K$. The server uses the keys stored at the users' caches to encode the delivery signal $X_{(d_1,\ldots,d_K)}$ such that the transmission is secure from the wiretapper. 
\begin{figure*}[t]
\centering \vspace{-5pt}
\subfigure[]{
\includegraphics[width=3.1 in,height=2.0 in]{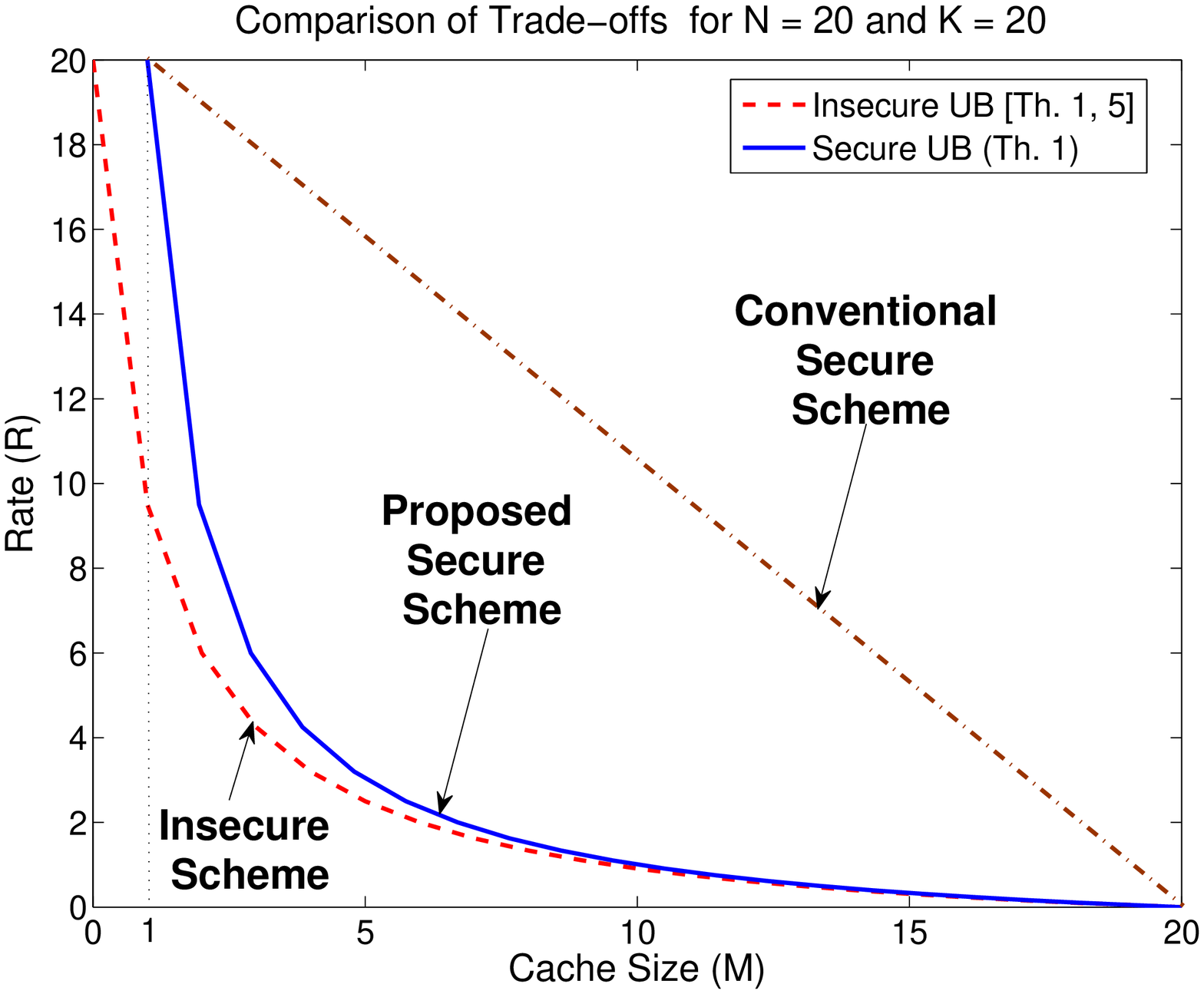}
\label{fig:n100k100}
}
\subfigure[]{
\includegraphics[width=3.2 in,height=2.1in]{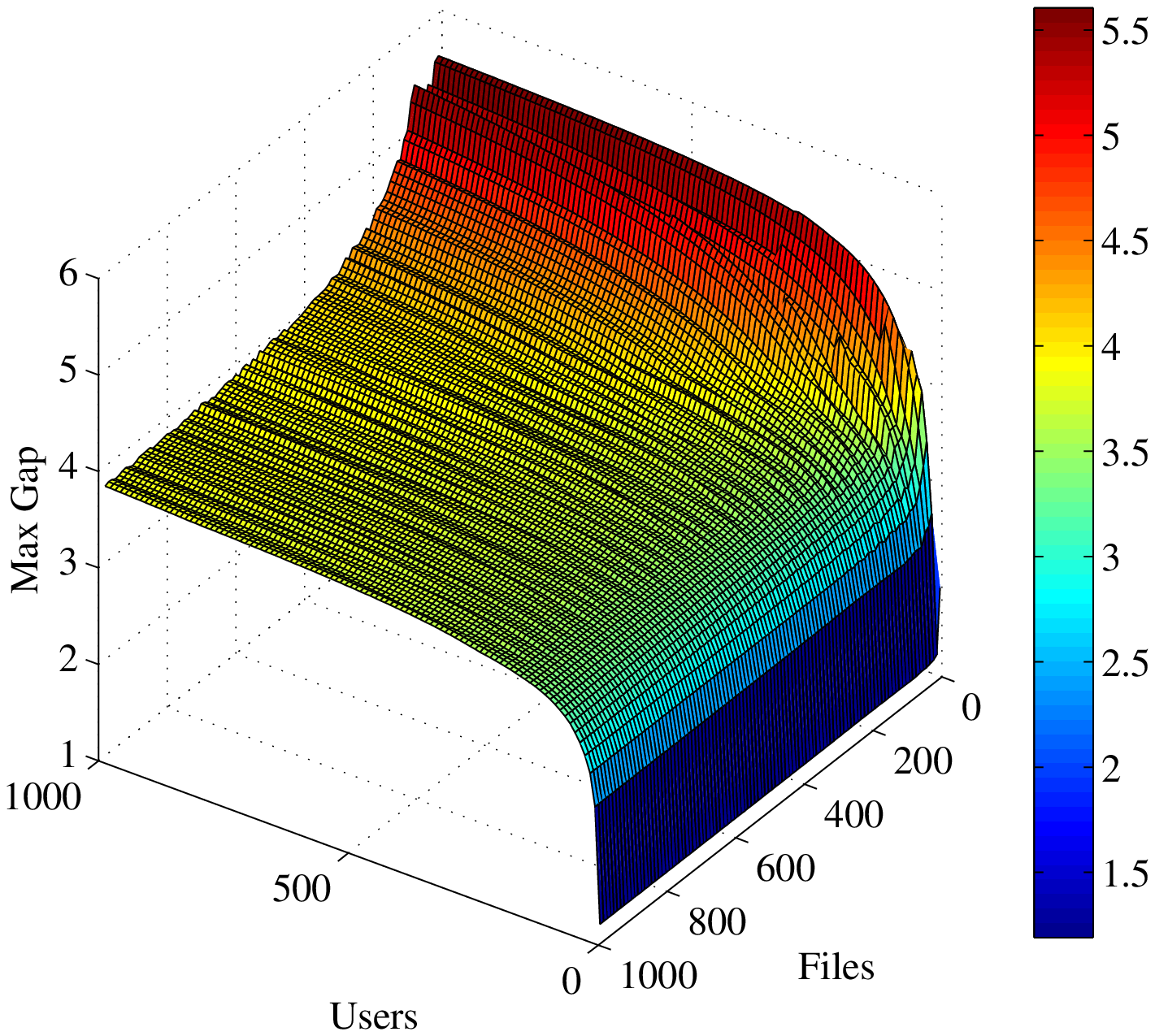}
\label{fig:mg}
}
\vspace{-6pt}
\caption{{(a)} Centralized Secure vs Non-Secure Bounds $N = K = 20$; {(b)} Multiplicative gap between $R_s^C(M)$ and lower bound on $R^*_s(M)$.} \vspace{-18pt}
\end{figure*}
\begin{thm}\label{th:1}
For \N~files and $K$ users, each with a cache size of $M \in \frac{(N-1)}{K} \cdot t + 1$, for $t \in \{0,1,2,\ldots,K\} $ we have \\
\begin{align}\label{eq:th2}
&R_s^*(M) \leq  R^C_s(M) \triangleq K\cdot\left(1-\frac{M-1}{N-1}\right) \left\{\frac{1}{1 + K\cdot\frac{M-1}{N-1}}\right\}
\end{align}
i.e., the rate $R^C_s(M)$ is securely achievable. For any $1\leq M\leq N$, the lower {convex} envelope of these points is achievable. 
\end{thm} \vspace{-5pt}
\noindent The algorithm achieving the rate in Theorem \ref{th:1} is presented in Algorithm \ref{alg:0} (Appendix \ref{ssec:th1}). Similar to \cite{Maddah-Ali}, the achievable rate in (\ref{eq:th2}) consists of three factors. The first factor $K$ is the worst case rate in the case when no data is cached $(M_D = 0)$. The second factor in (\ref{eq:th2}) is $\left(1 - \frac{M-1}{N-1}\right)$. This is the \textit{secure local caching gain} and is relevant whenever $M$ is of the order of $N$. The third factor in (\ref{eq:th2}) is $1/\left(1 + K \cdot\frac{M-1}{N-1}\right)$, which is the \textit{secure global caching gain}. Comparing Theorem \ref{th:1} to (Th.1, \cite{Maddah-Ali}), we observe that the terms $\frac{M}{N}$ in (Th.1, \cite{Maddah-Ali}) have been replaced by $\frac{M-1}{N-1}$. However, the combination of the global and local gains leads to the rate in (\ref{eq:th2}) being higher than the rate in (Th.1, \cite{Maddah-Ali}) for a given value of $M,N$. This is the cost paid for the security in the system. However, as $K,N$ become large, the secure rate is asymptotically equal to the non-secure case. When $N=K=20$, it can be seen from Fig. \ref{fig:n100k100} that the secure and non-secure bounds almost coincide i.e.,  security from a wiretapper can be achieved at \textit{almost negligible cost} for a large number of files and users.

Consider the case of conventional unicast content delivery to each user. In contrast to the insecure scheme in \cite{Maddah-Ali}, to make the delivery phase secure, however, each user has to store a unique key (of the same size as a single file). During delivery, the server encodes the user's requested file with its key and transmits it. Thus, even with no data storage in cache, the cache size has to be at least $F$ bits to store a key $(M_K = 1)$ i.e., in the secure problem, $M=0$ is infeasible. The worst case rate is achieved at $M=1$ and the $(M,R_s^C)$ pair $(1,K)$ is achievable. At the other extreme when $M = N$ i.e., the case where all files are stored in the user's cache and no content delivery is required. In this case $M_D = N, M_K = 0$ and the $(M,R_s^C)$ pair $(N,0)$ is achievable. We refer to a scheme which achieves points on the line joining $(1,K)$ and $(N,0)$ as the \textit{conventional secure scheme}, where each user stores one unique key and encrypted files are unicast to each user based on their request. On the other hand, the proposed scheme in Algorithm \ref{alg:0} jointly designs the placement of data and keys in the users' caches such that \textit{coded secure multicasting} can be achieved among users. Next, we present a lower bound on $R^*_s(M)$ stated in the following theorem. 

\vspace{-5pt}
\begin{figure*}[t]
\centering \vspace{-5pt}
\subfigure[]{
   \includegraphics[width=3.1 in,height=2.1 in]{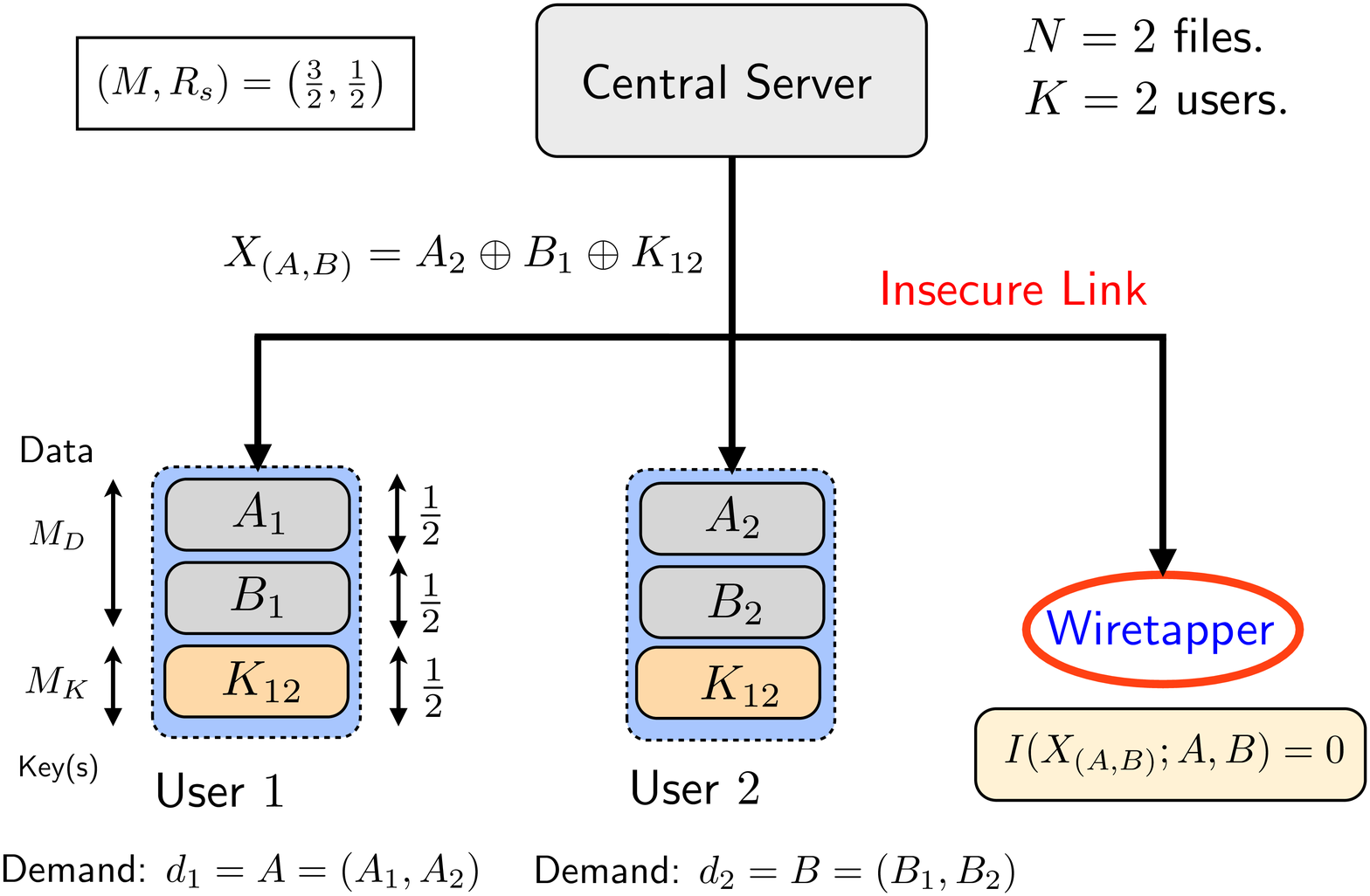}
   \label{fig:n2k2sys} 
 }
\subfigure[]{
   \includegraphics[width=3.1 in,height=2.1 in]{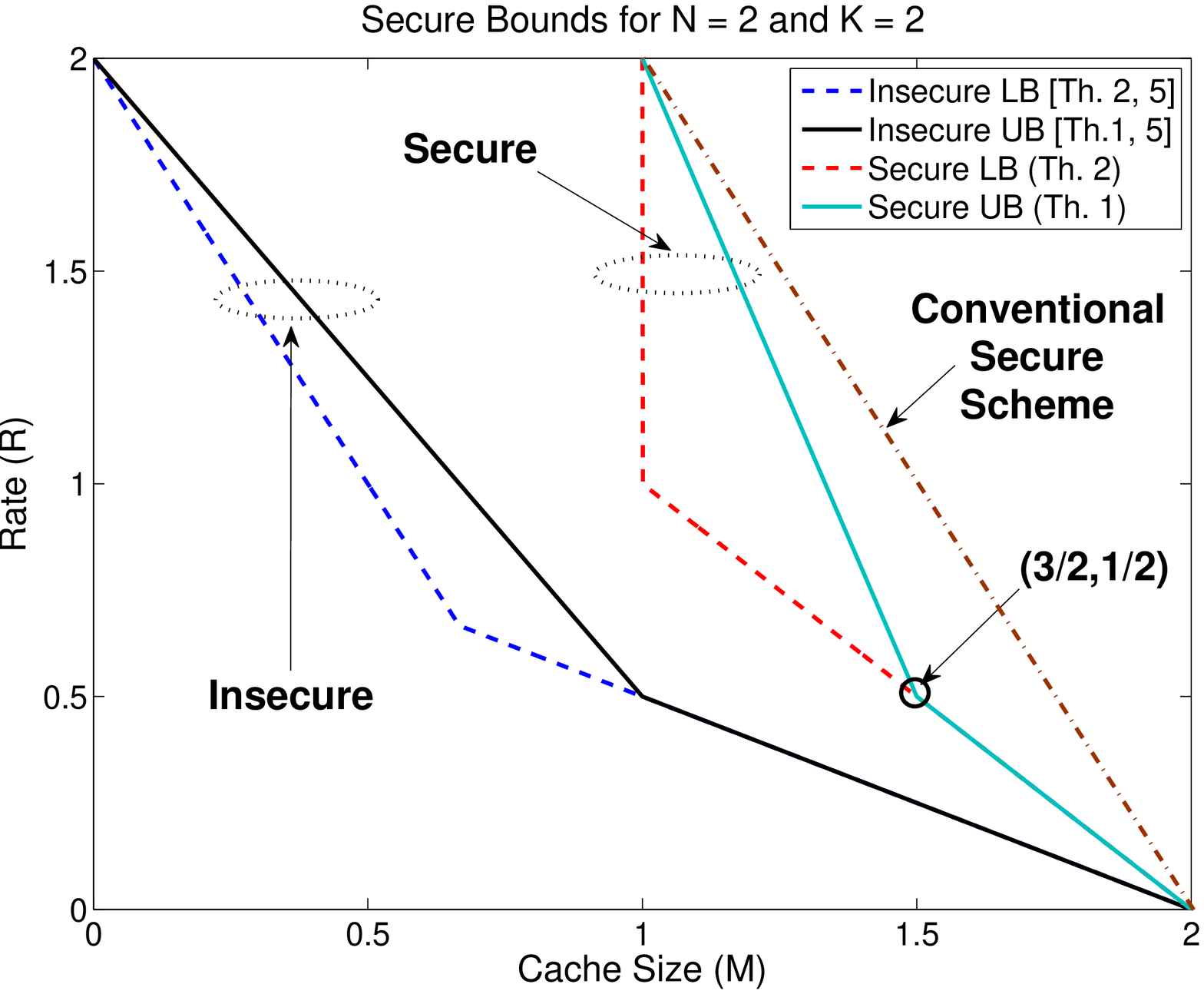}
   \label{fig:n2k2}
	 }
 	\vspace{-5pt}
\caption{{(a)}  Secure Caching Scheme and {(b)} $(M,R^C_s)$ trade-off for $N = K = 2$.}\vspace{-18pt}
\end{figure*}

\begin{thm}\label{th:3}
For $N$ files and $K$ users, each having a cache size $1\leq M \leq N$,
\begin{equation}\label{eq:th3}
R^*_s(M) \geq \max_{s\in \{1,\ldots,\min\{N,K\}\}} \left(s - \frac{s(M-1)}{\left(\lfloor \frac{N}{s} \rfloor - 1\right)}\right).
\end{equation}
\end{thm}
\noindent The proof of Theorem \ref{th:3} is presented in Appendix \ref{ssec:th3}. Next, we compare the achievable rate from Theorem \ref{th:1} and the lower bound  on the optimal rate in Theorem \ref{th:3}, and show that a constant multiplicative gap exists between $R_s^*(M)$ and the achievable rate $R^C_s(M)$. 
\begin{thm}\label{th:4}
For $N$ files and $K$ users, each having a cache size {$\max\left\{\frac{(K-N)(N-1)}{KN} + 1,1\right\}\leq M \leq N$,}
\begin{equation}\label{eq:th4}
1 \leq \frac{R^C_s(M)}{R^*_s(M)} \leq 17.
\end{equation}
\end{thm}
\noindent The proof of Theorem \ref{th:4} is presented in Appendix \ref{ssec:th4}. The gap is unbounded and scales with $K$ only for the case of $K>N$ in the regime $1\leq M < \frac{(K-N)(N-1)}{KN} + 1$, which is negligibly small for large $K,N$ as discussed in Appendix \ref{ssec:th4}. While the analytical constant of 17 is large for practical purposes, the gap can tightened numerically. Fig. \ref{fig:mg} shows the maximum value of the multiplicative gap between $R^C_s(M)$ and the lower bound on $R^*_s(M)$ for values for $N,K$ ranging from $1$ to $1000$ and all feasible values of $M$ in each case. It can be seen that the gap is generally less than $4$ when $K<N$. However for $K>N$, and for small $N$, the gap is larger i.e., around $6$. \vspace{-5pt}

\subsection{Intuition behind Theorem \ref{th:1} (Achievability)}
We next present a series of examples to explain the intuition behind the achievable rate in Theorem \ref{th:1} and highlight the interesting features of the proposed secure delivery scheme. 
\begin{exmp}\label{ex1}
We illustrate the achievable scheme in Theorem \ref{th:1} for the case of $N=2$ files and $K=2$ users. From Theorem \ref{th:1} we have $M\in \frac{2-1}{2}\{0,1,2\} + 1 = \{1,\frac{3}{2},2\}$ are the possible cache sizes for each user. Let the two files be $W_1 = A$ and $W_2 = B$. The bounds in Theorems \ref{th:1} and \ref{th:3} are shown in Fig. \ref{fig:n2k2} along with the bounds for the non-secure case from \cite{Maddah-Ali}. We start with the upper bound in Theorem \ref{th:1}. Considering the extreme point $M=1$, the cache of both users $Z_1,Z_2$ only stores two unique keys $\mathcal{K}_1 ,\mathcal{K}_2$ and the server transmits both the files $A,B$ over the shared link XOR-ed with a key. Given the worst-case demand $(d_1,d_2) = (A,B)$, the server can transmit $X_{(A,B)} = \{A\oplus \mathcal{K}_1, B\oplus\mathcal{K}_2\}$. This system satisfies every possible request with rate $R=2$ and it is easily verified that $I\left(X_{(A,B)};A,B\right) = 0$. Thus $(M,R^C_s) = (1,2)$ is \textit{securely} achievable. At the other extreme, when $M=2$, each user can cache both files and no transmission is necessary. Hence the $(M,R^C_s) = (2,0)$ is \textit{securely} achievable. 

Now we consider the intermediate case in which $M=3/2$. The scheme for this scenario is depicted in Fig. \ref{fig:n2k2sys}. Both the files are split into $2$ equal parts: $A=(A_1,A_2)$ and $B=(B_1,B_2)$, where $A_{1}, A_{2}, B_{1}, B_{2}$ are each of size $F/2$ bits. We also generate a key $\mathcal{K}_{12} \sim \textrm{unif}\{1,\ldots,2^{(F/2)}\}$, which is independent of both the files $A,B$ and has the same size as the sub-files i.e., $F/2$ bits. In the storage phase, the server fills the caches as follows: $Z_1 = (A_1,B_1,\mathcal{K}_{12})$ and $Z_2 = (A_2,B_2,\mathcal{K}_{12})$ i.e., each user stores one exclusive part of each file and the key. Thus $M_D = 1/2 + 1/2 = 1$ and $M_K = 1/2$. Now, consider the worst case request $(d_1,d_2) = (A,B)$. In order to satisfy this request, user 1 requires the file fragment $A_2$ while user 2 requires the file fragment $B_1$. In this case, the server transmits $X_{(A,B)} = \{A_2\oplus B_1\oplus \mathcal{K}_{12}\}$ which is of rate $1/2$. User 1 can obtain $A_2$ by XOR-ing out $B_1,\mathcal{K}_{12}$ while user 2 can get $B_1$ by XOR-ing out $A_2,\mathcal{K}_{12}$ from $X_{(A,B)}$. A wiretapper, on the other hand, would gain no knowledge of either file from the transmission since $I\left(X_{(A,B)};A,B\right)=0$ which follows from the fact that the key $\mathcal{K}_{12}$ is uniformly distributed. Thus, $(M,R^C_s) = ({3}/{2},{1}/{2})$ is \textit{securely} achievable. This can be seen in the secure upper bound in Fig. \ref{fig:n2k2}. 
Given that the points $(1,2)$,$({3}/{2},{1}/{2})$ and $(2,0)$ are achievable, the lines joining pairs of these points are also achievable. Thus, this proves the achievability of the secure upper bound in Fig \ref{fig:n2k2}. The gap between the insecure and secure achievable bounds results from the storage of the key in the users' cache. \hfill $\Diamond$
\end{exmp}
In the two user example, there is only a single key $\mathcal{K}_{12}$ in the system. Thus, if the key is compromised, the security of the entire system fails. The scheme proposed in Theorem \ref{th:1} for general values of $(N,K)$, however is more robust in its key management when the number of files and users increase. We next illustrate this point through an example. 
\begin{figure*}[t]
\centering \vspace{-5pt}
\subfigure[]{
   \includegraphics[width=3.1 in,height=2.1 in]{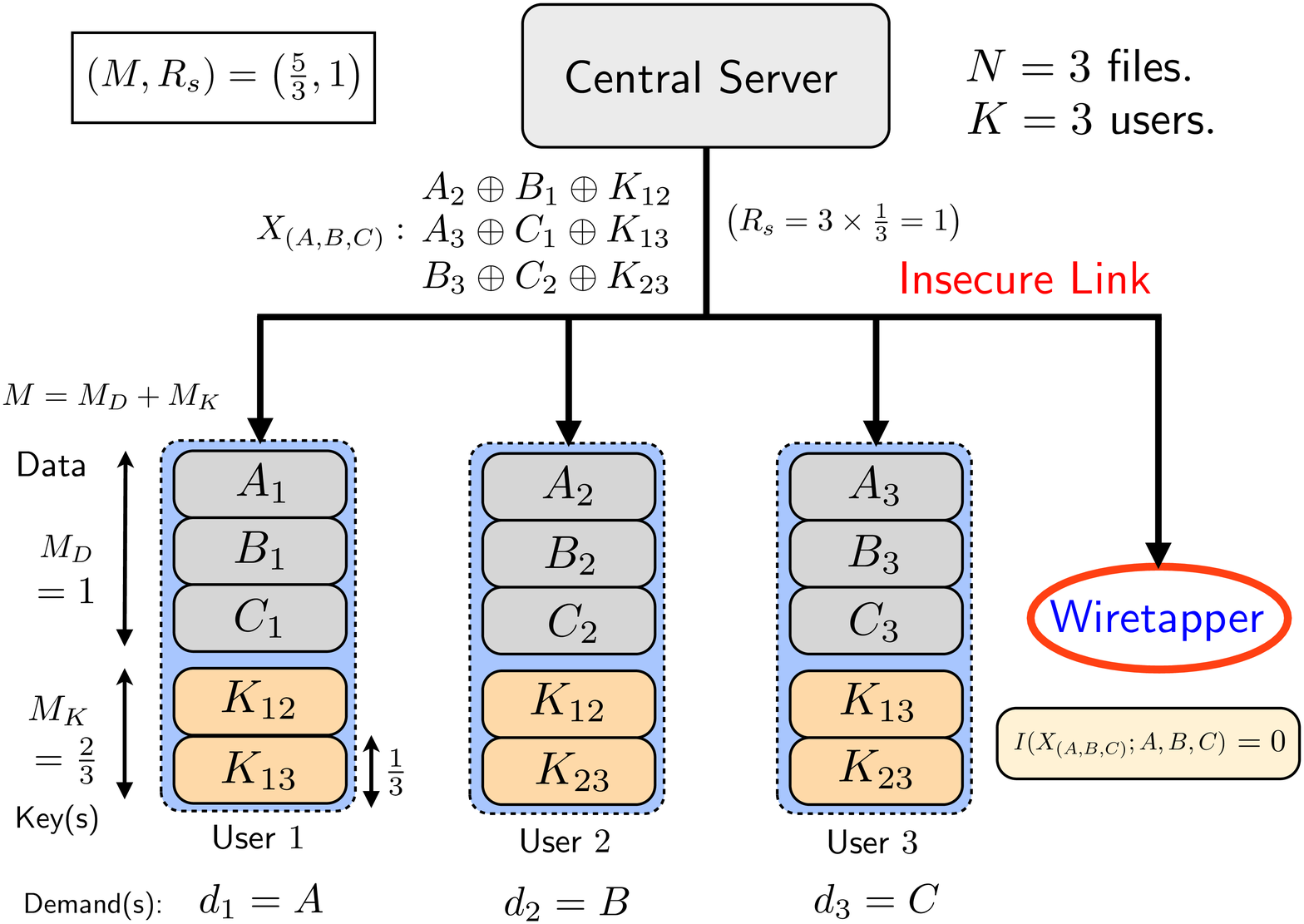}
   \label{fig:n3k3sys} 
 }
\subfigure[]{
   \includegraphics[width=3.1 in,height=2.1 in]{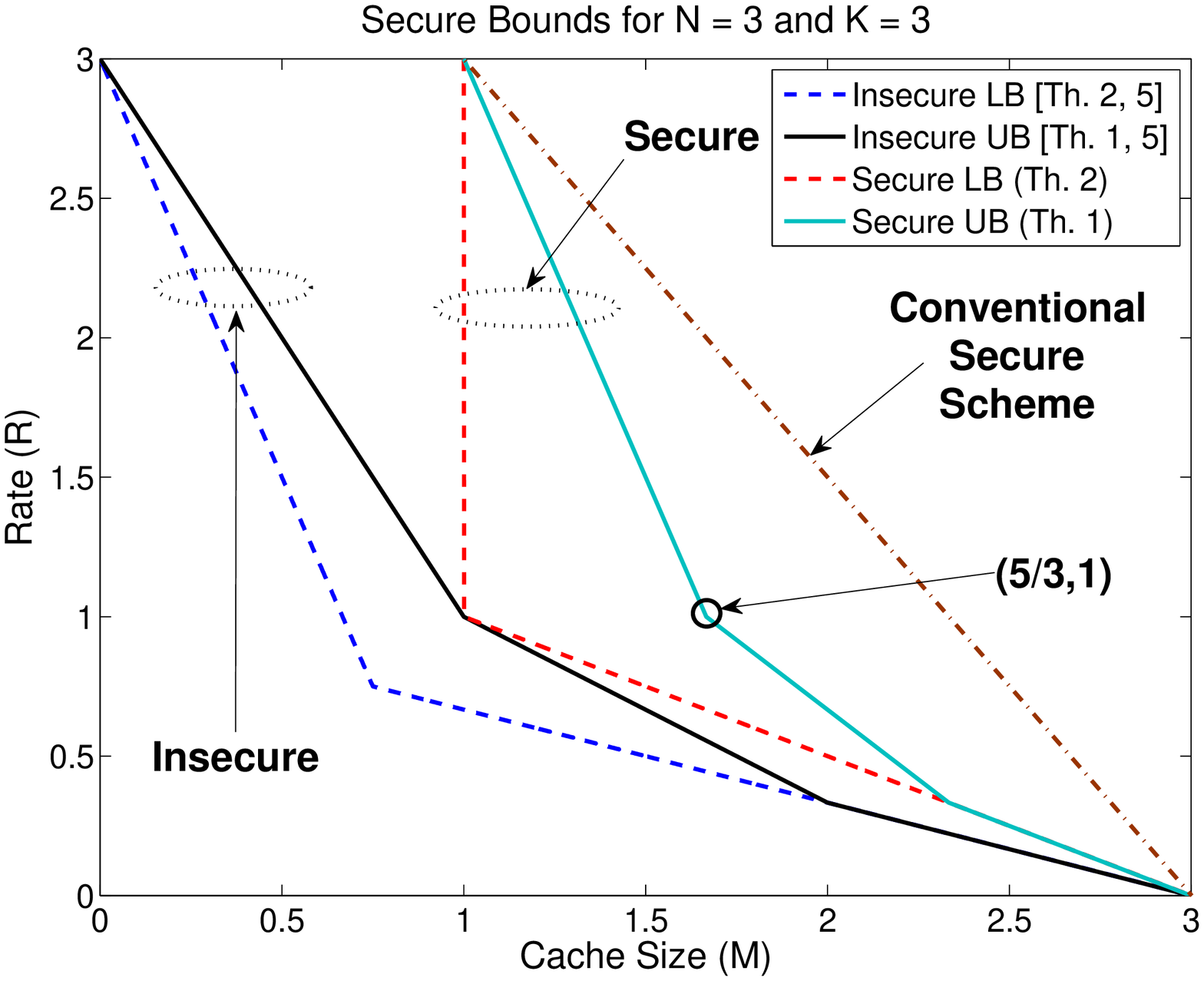}
   \label{fig:n3k3}
 } \vspace{-5pt}
\caption{{(a)} Secure Caching Scheme and {(b)} $(M,R^C_s)$ trade-off for $N = K = 3$.}\vspace{-18pt}
\end{figure*}
\begin{exmp}\label{ex2}
We consider the case for $N=K=3$. For this case, from Theorem \ref{th:1}, $M\in\{1,\frac{5}{3},\frac{7}{3},3\}$. The system and bounds for this case are illustrated in Fig. \ref{fig:n3k3sys} and \ref{fig:n3k3}. We consider the case of $M=5/3$ and three files $A,B,C$.  Each file is split into $3$ equal parts i.e., $A = (A_1,A_2,A_3),~B = (B_1,B_2,B_3),~C = (C_1,C_2,C_3)$. We also have $3$ keys in the system, $\mathcal{K}_{12},\mathcal{K}_{13},\mathcal{K}_{23}$. In this case, each subfile and each key is of size $F/3$ bits. In general, the key $\mathcal{K}_{ij}$ is placed in the caches of users $i$ and $j$. The keys are chosen combinatorially and a general strategy is discussed in Appendix \ref{ssec:th1}. The overall cache placement is as follows: $Z_1 = \{A_1,B_1,C_1, \mathcal{K}_{12},\mathcal{K}_{13}\}$, $Z_2 = \{A_2,B_2,C_2, \mathcal{K}_{12},\mathcal{K}_{23}\}$ and $ Z_3 = \{A_3,B_3,C_3, \mathcal{K}_{13},\mathcal{K}_{23}\}$.
Thus each cache has size $M = 5\times(1/3) = 5/3$, where $M_D = 1, M_K = 2/3$. Now considering a worst case request where all users request different files, $(d_1,d_2,d_3) = (A,B,C)$, the server can make the transmission, $X_{(A,B,C)} = \left\{\{A_2 \oplus B_1 \oplus \mathcal{K}_{12}\}, \{A_3 \oplus C_1 \oplus \mathcal{K}_{13}\}, \{B_3 \oplus C_2 \oplus \mathcal{K}_{23}\}\right\}$, such that everyone can securely retrieve their requested files.
Thus $(M,R^C_s)=(5/3,1)$ is \textit{securely} achievable since $I(X_{(A,B,C)}; A, B, C)=0$ i.e., a wiretapper would gain no information about the files from the transmission. It can be seen from the cache contents that there are multiple keys in the system thereby avoiding a single point of failure. In general, if we choose operating points $(M,R^C_s)$ such that $M_K > 1/K $, single points of failure in the system can be avoided. Thus the scheme forms an interesting memory-rate trade-off based on users' security constraints which is elaborated subsequently in Remark \ref{rm:1}. \hfill $\Diamond$
\end{exmp} 

\begin{remark}[\textit{Key Memory vs. Data Memory Trade-off}]\label{rm:1}
The trade-off between the fraction of cache memory occupied by the data and the keys in the secure caching system is shown in Fig. \ref{fig:key} for $N=5$ files and $K=5$ users. Consider the cache memory constraint in Theorem \ref{th:1} i.e., $M \in \frac{N-1}{K}t + 1,~\forall t \in \{0,1,2,\ldots,K\}$. Now, since $M = M_D+M_K$, from Appendix \ref{ssec:th1}, we have $M_K = 1 - t/K$ and $M_D = Nt/K$. From Fig. \ref{fig:key}, it can be seen that $M_K$ dominates at lower values of $M$. Formally, $M \geq 2N/(N+1)$, data memory dominates key memory i.e., $M_D>M_K$. From Appendix \ref{ssec:th1}, we have ${K\choose t+1}$ unique keys in the system. Thus the case for there being only one unique key in the system corresponds to $t=K-1$ i.e., $M_K = 1/K$. Thus for avoiding one shared key across all users i.e., a single point of failure in the system, we need $M_K>1/K \Rightarrow t \leq K-1$, which corresponds to $M \leq (N-1)(K-1)/K + 1$. It is also undesirable that new keys be redistributed to the entire system each time a user leaves. The proposed scheme avoids this scenario by sharing keys. In case a user leaves or is compromised, only the keys contained in that user's cache need to be replaced, leaving the others untouched. Thus, a desirable region of operation would be: 
\begin{align*}
\frac{2N}{(N+1)} \leq M \leq \frac{(N-1)(K-1)}{K} + 1. 
\end{align*}
In general, a close inspection of Algorithm \ref{alg:0} reveals that when { $t > (K-r)$ i.e., when $M > (N-1)(K-r)/K + 1$}, a wiretapper can obtain all the keys in the system if it gains access to any $r$ of the $K$ user caches. This means that if $r$ users are compromised, system security will be violated. It is a trivial fact that at $t = 0$, $M=1$ and each user has one unique key. In this case, the wiretapper will need access to all  caches in order to violate the security of the system. 
\begin{figure}[h] \vspace{-8pt}
\centering
\includegraphics[width=3.2 in,height=2.2 in]{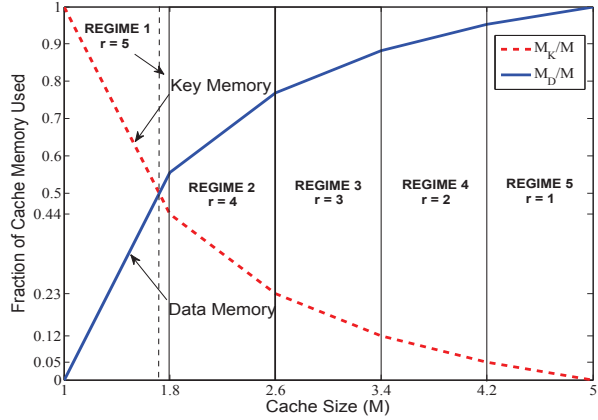}
\vspace{-8pt}
\caption{$M_K$ vs. $M_D$ tradeoff for $N=K=5$}\vspace{-5pt}\label{fig:key}
\end{figure}

From Fig. \ref{fig:key}, we can see that Regime 5, i.e., when $r=1$, is the weakest regime from the security perspective as there is only one key in the system. Thus operation in Regimes 1-4 is desirable for the case of $N=K=5$. Now, considering the \textit{conventional secure scheme}, it is seen that there is no sharing of keys as each transmission is useful to only one user. Thus each user stores an unique key of size $|\K| = (1-\frac{M-1}{N-1})F$ bits. This scheme thus requires the wiretapper to have access to all the caches for the system security to be compromised. Comparing the conventional and proposed schemes from a security perspective, we see that the proposed scheme is a \textit{trade-off} between security and minimization of the rate over the shared link. While the conventional scheme is more difficult to compromise for $M \in \mathbb{N}$, the proposed scheme is able to improve on the transmission rate significantly while still providing security. \hfill $\Diamond$
\end{remark} \vspace{-5pt}
\subsection{Intuition behind Theorem \ref{th:3} (Converse)}
We next present the main idea behind the proof of the converse stated in Theorem \ref{th:3} through a novel extension of the cut-set bound to incorporate the security constraint. To this end, we focus on the caching system with $N=2$ files (denoted by $A$ and $B$) and $K=2$ users (with cache contents denoted by $Z_{1}$ and $Z_{2}$). Consider the scenario where user $1$ demands file $A$ and user $2$ demands file $B$, i.e., the demand vector is $(d_{1}, d_{2})=(A, B)$.  
It is easy to check that using the communication $X_{(A,B)}$ from the central server along with the two caches $Z_1,Z_2$, both files $(A, B)$ can be recovered. This implies the following constraint: 
\begin{align}
& H\left(A,B|X_{(A,B)},Z_1,Z_2\right) \leq \epsilon. \label{exp:FR1}
\end{align}
Next, for the communication $X_{(A,B)}$ to be secure, we also require the following security constraint to hold: 
\begin{align}
& I\left(A,B;X_{(A,B)}\right) \leq \epsilon. \label{exp:sec1}
\end{align}
Using these two constraints, we next show that for any scheme, $M\geq 1$ must necessarily hold. From the constraints (\ref{exp:FR1})-(\ref{exp:sec1}), we have the following sequence of inequalities:
\begin{eqnarray*}
2F & \leq & H(A,B) = I\left(A,B;X_{(A,B)},Z_1,Z_2\right) + {H\left(A,B|X_{(A,B)},Z_1,Z_2\right)}\nonumber\\
	 & \myleq{(\ref{exp:FR1})} &	I\left(A,B;X_{(A,B)},Z_1,Z_2\right) + \epsilon  \nonumber\\
	 & = & {I\left(A,B;X_{(A,B)}\right)} + I\left(A,B;Z_1,Z_2|X_{(A,B)}\right) + \epsilon \nonumber\\
	 & \myleq{(\ref{exp:sec1})} & I\left(A,B;Z_1,Z_2|X_{(A,B)}\right) + 2\epsilon  \nonumber\\
	 & \leq & H\left(Z_1,Z_2|X_{(A,B)}\right)  + 2\epsilon \leq H(Z_1,Z_2) + 2\epsilon \nonumber\\
	 & \leq &  H(Z_1) + H(Z_2)  + 2\epsilon  \leq 2MF  + 2\epsilon. 
\end{eqnarray*}
This implies
\begin{equation}
M  \geq  1 - \frac{\epsilon}{F}.	\label{eq:t1l2}
\end{equation}
Taking the limit $\epsilon \rightarrow 0$, we arrive at the proof of $M\geq 1$.
Now consider the fact that given the transmissions from the server $X_{(A,B)}$ for demands $(d_1,d_2) = (A,B)$, $X_{(B,A)}$ for demands $(d_1,d_2) = (B,A)$ and one cache $Z_1$, both the files $A,B$ can be recovered. Again, we have the following constraints for file retrieval and security: 
\begin{align}
& H\left(A,B|X_{(A,B)},X_{(B,A)},Z_1\right) \leq \epsilon  \label{exp:FR2} \\
& I\left(A,B;X_{(A,B)}\right) \leq \epsilon. \label{exp:sec2}
\end{align}
Thus we have, 
\begin{eqnarray*}
2F & \leq & H(A,B) =  I\left(A,B;X_{(A,B)},X_{(B,A)},Z_1\right) +  {H\left(A,B|X_{(A,B)},X_{(B,A)},Z_1\right)}\nonumber\\
	 & \myleq{(\ref{exp:FR2})}&  I\left(A,B;X_{(A,B)},X_{(B,A)},Z_1\right) + \epsilon\nonumber\\ 
	 & = & {I\left(A,B;X_{(A,B)}\right)} + I\left(A,B;X_{(B,A)},Z_1|X_{(A,B)}\right) + \epsilon\nonumber\\
	 & \myleq{(\ref{exp:sec2})}& I\left(A,B;X_{(B,A)},Z_1|X_{(A,B)}\right) + 2\epsilon\nonumber\\
	 & \leq & H\left(X_{(B,A)},Z_1|X_{(A,B)}\right) + 2\epsilon\nonumber\\
	 & \leq & H\left(X_{(B,A)}\right) + H(Z_1) + 2\epsilon\nonumber\\
	 & \leq & R^*_sF + MF + 2\epsilon.
\end{eqnarray*}
This implies that
\begin{equation}\label{eq:t2l1}
R^*_s  + M  \geq  2 - \frac{2\epsilon}{F}.	
\end{equation}
Taking the limit $\epsilon \rightarrow 0$, we arrive at the proof of $R^*_s  + M  \geq  2$. We can see that both (\ref{eq:t1l2}) and (\ref{eq:t2l1}) hold for all achievable $(M,R_s)$ pairs. Thus we have, $R^*_s(M) \geq 2 - M$ and $M \geq 1$ which gives the lower bound in Fig. \ref{fig:n2k2}.

\begin{figure*}[t]
\centering \vspace{-5pt}
\subfigure[]{
   \includegraphics[width=3.2 in,height=2.1 in]{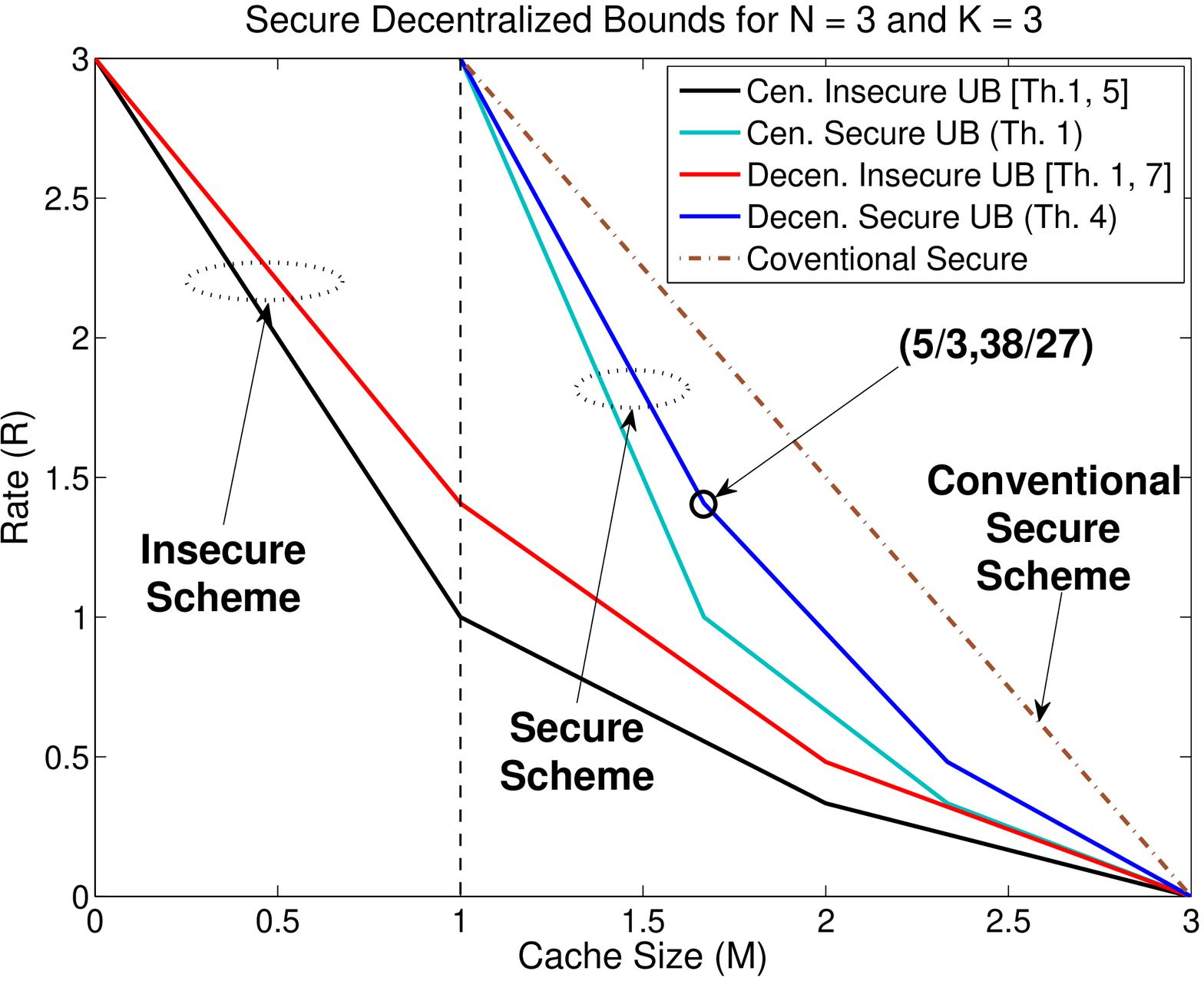}
   \label{fig:dec_n3k3}
	 }
	\subfigure[]{
   	\includegraphics[width=3.1 in,height=2.1 in]{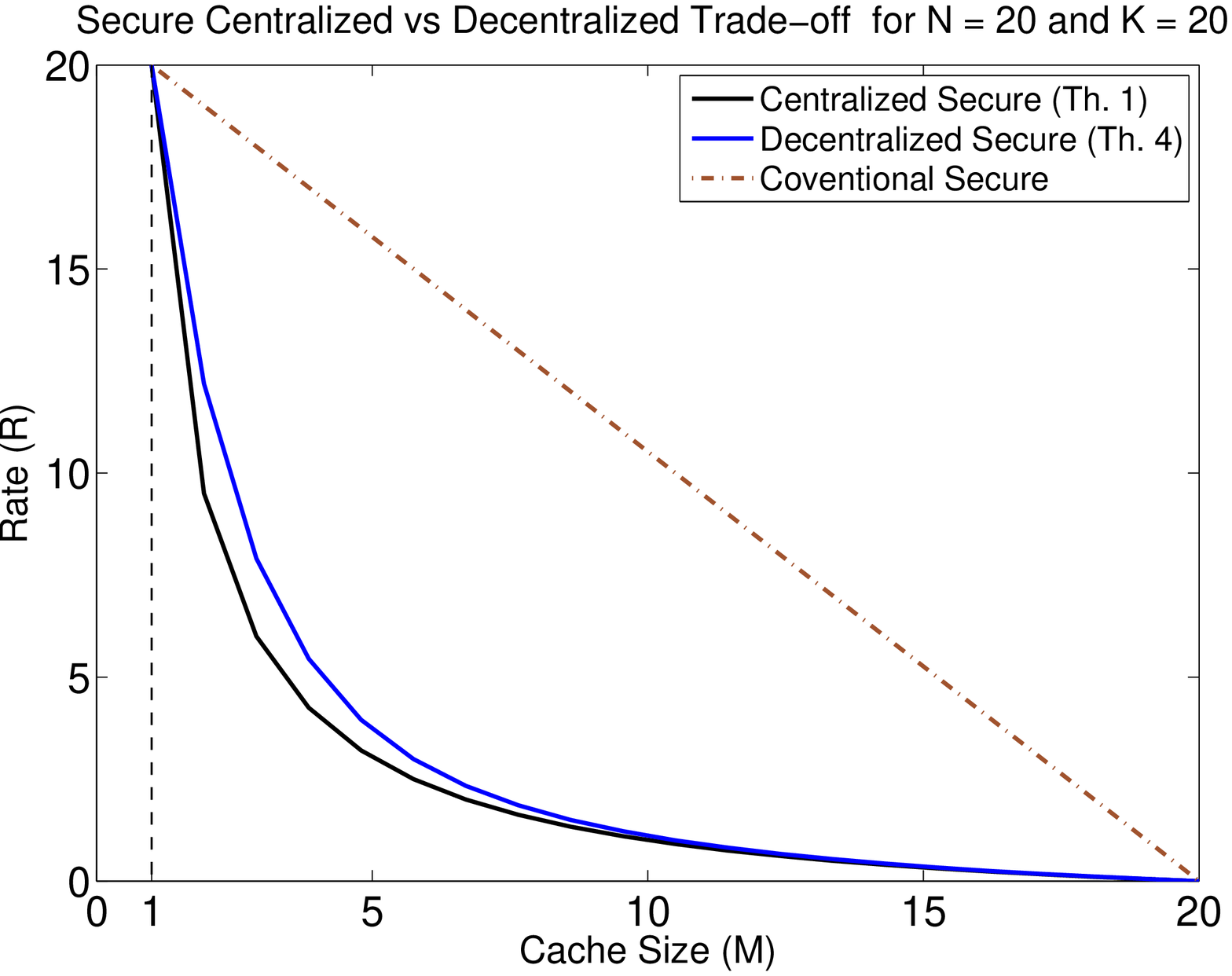}
	\label{fig:cen_dec}
	 }\vspace{-6pt}
\caption{{(a)} $(M,R^D_s)$ trade-off for $N=K=3$ and {(b)} Centralized vs Decentralized Secure Bounds for $N = K = 20$.} \vspace{-18pt}
\end{figure*}
\vspace{-5pt}
\section{Decentralized Caching with Secure Delivery}
\label{sec:mainres2}
In this section, we extend the secure caching problem to a decentralized caching scheme as discussed in \cite{Maddah-Ali-decentralized}. In the decentralized caching scheme, each user is allowed to cache any random $\frac{M-1}{N-1}$ bits of each of the $N$ files in the system. In the coded delivery scheme, the central server maps the contents of individual users' caches to fragments (which contain non-overlapping combination of bits) in each file. The fragments reflect which user (or set of users) has cached bits contained in the given fragment. This phase is followed by a centralized key placement procedure where the server stores shared keys in each user's cache. The key placement needs to be centralized to maintain key integrity and to secure the files from an external wiretapper. In the delivery phase, the server receives a request $(d_1,\ldots,d_K)$ and forms coded multicast transmissions to extract global caching gain from the system. It then encodes the transmissions with the shared keys and transmits them over the multicast link. The decentralized algorithm is presented in Algorithm \ref{alg:1} in Appendix \ref{ssec:th5}.

In the case of decentralized caching, similar to the centralized case, the\textit{ conventional secure scheme} is one which stores only one unique key per user and exploits only the local caching gain by using encrypted unicast delivery. The transmission rate in this case is given by $K(1- \frac{M-1}{N-1})$. After the cache placement, the server chooses the scheme which provides the minimum rate over the shared link. The secure rate is then characterized by the following theorem.
\begin{thm}\label{th:5}
For $N$ files and $K$ users, each with a cache size of $M \in \frac{N-1}{N}\cdot t + 1$, for $t\in(0,N]$,
\setcounter{equation}{16}
{\begin{align}\label{eq:th5}
& R^D_s(M) \triangleq K\left(1 - \frac{M-1}{N-1}\right)\cdot\min\left\{\frac{N-1}{K(M-1)}\cdot\left(1-\left(1 - \frac{M-1}{N-1}\right)^K\right), 1\right\}
\end{align}}
is securely achievable. For any {$1< M \leq N$}, the lower {convex }envelope of these points is achievable. 
\end{thm}
\noindent The proof of Theorem \ref{th:5} is given in Appendix \ref{ssec:th5}. The variable $t = M_D$, represents the part of the cache memory used to store data at each user (as detailed in Appendix \ref{ssec:th5}). Theorem \ref{th:5} is defined for $t>0$. At $t=0$, $M=1$ i.e., the caches store a single key of the size of each file. Entire files, XOR-ed with the keys, are then transmitted over the shared link. Thus the rate in this case is $R^D_s(1) \triangleq K$. As before, the same argument holds for the infeasibility of the secure scheme for $M=0$. The following example illustrates the caching scheme which achieves the rate in Theorem \ref{th:5}.

\begin{exmp}
We consider the case for $N=3$ files and $K=3$ users, each with a cache of size $MF$ bits. Let the three files be denoted as $(W_1,W_2,W_3) = (A,B,C)$. Fig. \ref{fig:dec_n3k3} shows the rate achieved by the secure decentralized caching scheme given by Theorem \ref{th:5}, the rate of the insecure decentralized scheme from \cite{Maddah-Ali-decentralized} and the corresponding centralized bounds. In the decentralized placement phase, each of the $3$ users caches a subset of $(M-1)F/2$ bits of each file independently at random. Thus, each bit of a file is cached by a specific user with probability $(M-1)/2$. Considering the file $A$, the server maps the storage of fragments of file $A$ at the different users' caches into splits, $A_{\mathcal{T}}$, such that $\mathcal{T} \subseteq \{1,2,3\},~ |\mathcal{T}|=i $ for $i = 0,1,2,3$. Thus there are $\sum_{i=0}^{3}{3 \choose i} = 2^3 = 8$ splits of file $A$: $(A_{\phi},A_{1},A_{2},A_{3},A_{12},A_{13},A_{23},A_{123})$, where $A_{\phi}$ consists of bits of $A$ which are not stored in any users' cache. On the other hand, $A_{123}$ has bits which are stored in all users cache. In general, bits in $A_{\mathcal{T}}$ are stored in user $k$'s cache if $k\in \mathcal{T}$. By law of large numbers, we have:
\begin{equation}\label{atau}
|A_{\mathcal{T}}| \approx \left( \frac{M-1}{2}  \right)^{|\mathcal{T}|}\left(1 - \frac{M-1}{2} \right)^{3 - |\mathcal{T}|}F ~~\textrm{bits}
\end{equation}
with probability approaching one for large enough file size F. The same analysis holds for files $B,C$. Next, we consider the generation of keys $\K_{\mathcal{S}}$ for $\mathcal{S}\subseteq \{1,2,3\}, |\mathcal{S}| = j$ for $j=1,2,3$. Thus the keys generated in the system are: $\K_{1},\K_{2},\K_{3},\K_{12},\K_{13},\K_{23},\K_{123}$. It can be seen that there are $2^K - 1 = 7$ unique keys in the system. Next we look at the cache contents from the central server's perspective after the centralized key placement phase and before the delivery procedure begins. The cache placement for $N=K=3$ is given in (\ref{cachepl}). 
\begin{figure*}[!b]\vspace{-10pt}
\setcounter{equation}{18}\small
\hrulefill \vspace{2pt}
\begin{equation}
\begin{array}{ccc}
	Z_1	= \left\{\begin{array}{c} A_1, A_{12}, A_{13},A_{123}\\
															 B_1, B_{12}, B_{13},B_{123}\\
													     C_1, C_{12}, C_{13},C_{123} \\
															 \K_1, \K_{12}, \K_{13}, \K_{123}\\	
																				\end{array} \right\} & Z_2	= \left\{\begin{array}{c} A_2, A_{12}, A_{23},A_{123}\\
																																														B_2, B_{12}, B_{23},B_{123}\\
																																														C_2, C_{12}, C_{23},C_{123} \\
																																													\K_2, \K_{12}, \K_{23}, \K_{123}\\
																																																	\end{array} \right\}  & Z_3	= \left\{\begin{array}{c} A_3, A_{13}, A_{23},A_{123}\\
																																																																												B_3, B_{13}, B_{23},B_{123}\\
																																																																												C_3, C_{13}, C_{23},C_{123} \\
																																																																											\K_3, \K_{13}, \K_{23}, \K_{123}\\
																																																																															 \end{array} \right\}.
\end{array}
\label{cachepl}																				                            
\end{equation}
\end{figure*}
The cache placement phase is entirely decentralized as the users do not need to consider the number of other users in the system or their cache contents while storing file fragments in their caches. Next, we consider the delivery procedure of the decentralized caching scheme. The system is characterized based on the worst possible rate over the shared link. Thus we consider a request $(W_{d_1},W_{d_2},W_{d_3}) = (A,B,C)$. The server responds by transmitting the reply $X_{(A,B,C)}$. Let the set $\mathcal{S} \subseteq \{1,2,3\}:|\mathcal{S}| = s$ for $s = 3,2,1$. Then we have $X_{(A,B,C)} = \left\{\mathcal{K}_{\mathcal{S}}\oplus_{k\in\mathcal{S}} W_{d_k,\mathcal{S}\setminus\{k\}} : k=1,2,3\right\}_{s=1}^{3},$ where $W_{d_k,\mathcal{S}\setminus\{k\}}$ corresponds to the fraction of the file $W_{d_k}$, requested by user $k$ which is not present in user $k$'s cache but is present in the cache of the other $s-1$ users in $\mathcal{S}$. Thus, for $K=3$ users in the system, the coded secure multicast delivery procedure has $3$ phases for each of $s = 3,2,1$. 

\noindent \underline{For $s=3~$:} We have $|\mathcal{S}| = 3 \Rightarrow \mathcal{S} = \{1,2,3\}$ and $|\mathcal{S}\setminus\{k\}| = 2$. The transmission is $\{A_{23} \oplus B_{13} \oplus C_{12} \oplus \K_{123}\}$. It can be seen that $\K_{123}$ is associated with sub-files $A_{23}, B_{13}, C_{12}$. Thus the size of the key is $|\K_{123}| = \max\{|A_{23}|, |B_{13}|, |C_{12}|\}$. In this case, each sub-file is zero padded to the size of the largest sub-file in the set. Considering user 1, we see that $Z_1$ contains $B_{13},C_{12}$ and $\K_{123}$. Thus user 1 can XOR out $A_{23}$ from the transmission. It can be seen that the same holds for users 2 and 3. Thus the transmission is useful for all users and the key makes it secure from the wiretapper. For $s=3$, there is only one transmission of the size of each of these sub-files. Thus, using (\ref{atau}), the rate over the shared link for this transmission is:
\begin{align}\label{er1}
\setcounter{equation}{19}
\left(\frac{M-1}{2} \right)^{2}\left(1 - \frac{M-1}{2} \right)F. 
\end{align}
\underline{For $s=2~$:}
 We have $|\mathcal{S}| = 2 \Rightarrow \mathcal{S} \in \{1,2\},\{2,3\},\{1,3\}$ and $|\mathcal{S}\setminus\{k\}| = 1$. The transmission for each subset $\mathcal{S}$ is $\left\{\{A_{2} \oplus B_{1} \oplus \K_{12}\}, \{B_{3} \oplus C_{2} \oplus \K_{23}\}, \{A_{3} \oplus C_{1} \oplus \K_{13}\}\right\}$. Again for user 1, we can see that $Z_1$ contains $B_1,C_1,\K_{12},\K_{13}$. Thus it can extract $A_2,A_3$ from this transmission. Similarly the other users can extract fragments of their requested files. In this case, there are three transmissions, each of the size of file fragment, say, $A_2$. Thus the rate of this transmission is:
\begin{align}\label{er2}
3\cdot\left(\frac{M-1}{2} \right)\left(1 - \frac{M-1}{2} \right)^{2} F.
\end{align}
\underline{For $s=1~$:} 
We have $|\mathcal{S}| = 1 \Rightarrow \mathcal{S} \in \{1\},\{2\},\{3\}$ and $|\mathcal{S}\setminus\{k\}| = 0$. The transmission for each subset $\mathcal{S}$ is $\left\{\{A_{\phi} \oplus \K_{1}\}, \{B_{\phi} \oplus \K_{2}\}, \{C_{\phi} \oplus \K_{3}\}\right\}$. These transmissions are sent to individual users, containing the residual fragments not stored in each user. The size of each transmission is equal to the size of the file fragments $A_{\phi},B_{\phi},C_{\phi}$. Thus the rate of this transmission is:
\begin{align}\label{er3}
3\cdot\left(1 - \frac{M-1}{2} \right)^{3} F.
\end{align}
Again considering user 1, we can see that the fragments of $A$ not present in its cache i.e., $A_{\phi},A_2,A_3,A_{23}$ are extracted from the entire transmission. The same holds true for the other users. The rate for the composite transmission $X_{(A,B,C)}$ is obtained by summing (\ref{er1}), (\ref{er2}) and (\ref{er3}):
\begin{align}
 & R^D_s(M)F = F\left(\frac{M-1}{2} \right)^{2}\left(1 - \frac{M-1}{2} \right) + 3F\left(\frac{M-1}{2} \right)\cdot\left(1 - \frac{M-1}{2} \right)^{2}   +  3F\left(1 - \frac{M-1}{2} \right)^{3} \nonumber\\
				  & = 3\left(1 - \frac{M-1}{2} \right)\frac{2}{3(M-1)}\left(1-\left(1 - \frac{M-1}{2}\right)^3\right)F,
\end{align}
which is the expression given in Theorem \ref{th:5} for $N=K=3$. Now, we have $M \in \frac{N-1}{N}\{1,2,\ldots,N\} + 1 = \left\{\frac{5}{3},\frac{7}{3},3\right\}$. Considering the point $M = {5}/{3}$, we have $R^D_s(M) = {38}/{27}$. Thus the pair $(M, R^D_s) = (5/3,38/27)$, is securely achievable. This is seen from the $(M,R^D_s)$ trade-off in Fig. \ref{fig:dec_n3k3}. Similarly other points on the trade-off curve can be evaluated using other feasible values of $M$. All points on the lines joining the achievable $(M,R_s^D)$ points are also achievable. \hfill $\Diamond$ 
\end{exmp}

Next, we consider the centralized and decentralized trade-off for a large number of files and users. Fig. \ref{fig:cen_dec} illustrates the case for $N=K=20$. Compared to Fig. \ref{fig:dec_n3k3}, we can see that as the number of files and users increase, the decentralized scheme approaches the centralized caching. Thus for large number of files and users, the rates are \textit{asymptotically equal}. This also implies that in the decentralized case, similar to the centralized case, that the cost for security is \textit{almost negligible} when number of files and users increase \cite{Secure-AR}. The following theorem and corollary compares the rate of the achievable secure decentralized scheme given in Theorem \ref{th:5} to the lower bound on the rate of the optimal secure scheme given in Theorem \ref{th:3} and the rate of the achievable secure centralized caching scheme given in Theorem \ref{th:1}.

\begin{thm}\label{th:6}
 Given $R^D_s(M)$ be the rate of the secure decentralized caching scheme given by Algorithm \ref{alg:1} and $R^*_s(M)$ be the rate of the optimal secure caching scheme, for $N$ files and $K$ users, each having a cache size {$\frac{N-1}{N} + 1\leq M \leq N$},
\begin{equation}\label{eq:th6}
\frac{R^D_s(M)}{R^*_s(M)} \leq 17.
\end{equation}
\end{thm}
\noindent The proof sketch of Theorem \ref{th:6} is given in Appendix \ref{ssec:th6}. Theorem \ref{th:6} implies that no scheme, regardless of complexity can improve by more than a constant factor upon the secure decentralized caching scheme presented in Algorithm \ref{alg:1} for the given regime of $M$. {The gap is unbounded only for the case of $K>N$ in the regime $1\leq M \leq \frac{N-1}{N} + 1$, which is negligibly small for large $N,K$ as discussed in Appendix \ref{ssec:th6}.}
\begin{corr}\label{cor3}
Let $R^C_s(M)$ be the rate of the secure centralized caching scheme given in Theorem \ref{th:1} and $R_s^D(M)$ be the rate of the secure decentralized caching scheme given in Theorem \ref{th:5}. For $N$ files and $K$ users, for {$\frac{N-1}{N} + 1\leq M \leq N$}, we have
\begin{equation}\label{eq:cor} 
\frac{R_s^D(M)}{R^C_s(M)} \leq 17.
\end{equation}
\end{corr}
\noindent Corollary \ref{cor3} is a direct outcome of Theorems \ref{th:4} and \ref{th:6}. It shows that the decentralized scheme is at most a constant factor $17$ worse than the secure centralized scheme in the given regime of $M$.
 
\vspace{-5pt}
\section{Discussion and Open Problems}
In this section, we discuss some of the open problems and extensions of the current work:

\noindent $\bullet$ \textit{Extension to Non-Uniform File Popularities and Multiple Demands per User:} The problem of caching with secure delivery discussed in this paper assumes all files have uniform popularity. We presented an extension of the secure delivery scheme to the case for non-uniform file popularities in \cite{aviksg-gws}. Furthermore, in this paper, we consider the secure caching problem for the case of single requests from users at a given time instant. However, an interesting case is when users demand multiple, say $L$, files at a given instant. The non-secure problem was addressed from an graph based index coding perspective in \cite{Ldem}, while for the secure case, it is an interesting area for future work.

\noindent $\bullet$ \textit{Noisy Links \& Multiple Eavesdroppers:} In the current treatment of the security problem, it is also interesting to note that the presence of multiple eavesdroppers would not alter the presented results since each eavesdropper would view the same multicast transmission which leaks no information about the files. This is due to the fact that we consider noiseless delivery in this model. The analysis of the problem for multiple eavesdroppers in the presence of noisy links is a direction of future research. 

\noindent $\bullet$ \textit{Extension to Multiple Requests over time:} Another area for future work is the case of security in delivering content for multiple requests over time i.e., security for an online coded caching scheme similar to the one in \cite{Maddah-Ali-Online} which would require a key generation technique such that collection of keys over time by an eavesdropper cannot lead to information leakage. 

\noindent $\bullet$ \textit{Closing the Gap in Small Buffer Case:} Finally closing the gap between the achievable rate and the information theoretic optimal secure rate for $K>N$ in the regime $1< M < \frac{(K-N)(N-1)}{KN} + 1$ for the centralized scheme and $1< M < \frac{N-1}{N} + 1$ for the decentralized scheme, is an interesting open problem.
\vspace{-2pt}\section{Conclusion}
\label{sec:conc}
In this paper, we have analyzed the problem of \textit{secure} caching in the presence of an external wiretapper for both \textit{centralized} and \textit{decentralized} cache placement. We have proposed a key based secure caching strategy which is robust to compromise of users and keys. We have approximated the information theoretic optimal rate of the secure caching problem with novel upper and lower bounds. It has been shown that there is a constant multiplicative gap between the optimal and the achievable rates for the given scheme in case of both centralized and decentralized caching scenarios for most parameters of practical interest. We have shown that for large number of files and users, the secure bounds approach that of the non-secure case i.e., the cost of security in the system is negligible when the number of files and users increase.


\appendices
\section{Proof of Theorem \ref{th:1}}\label{ssec:th1}
In this section, we discuss the secure centralized caching strategy which achieves the upper bound $R^C_s(M)$ as stated in Theorem \ref{th:1}. The algorithm achieving the rate in Theorem \ref{th:1} is presented in Algorithm \ref{alg:0}. These are two phases in the caching strategy: the storage phase and the delivery phase. We consider a cache size $M \leq N$ and $M \in \frac{N-1}{K} \cdot \{0,1.\ldots,K\} + 1$. Let $t \in \{0,1,\ldots,K\}$ be an integer between $0$ and $K$. The cache memory size can then be parametrized by $t$ as:
\begin{equation}\label{cache}
M = \frac{N-1}{K}t + 1 = \frac{Nt}{K} + 1 - \frac{t}{K}. 
\end{equation}
From (\ref{cache}), we have $t = \frac{K(M-1)}{N-1}$. Next, we break up the total cache memory into data memory and key memory, $M = M_D + M_K$, as follows:
\begin{eqnarray}
M_K  = 1 - \frac{t}{K}; ~~ M_D = M - M_K = \frac{Nt}{K}.
\end{eqnarray} 
From the discussion in Section \ref{sec:mainres}, we know that the \textit{conventional secure scheme} achieves the $(M,R_s^C)$ pair $(1,K)$ and $(N,0)$. Thus $R^*_s(1) \leq K$ and $R^*_s(N) = 0$. We therefore consider the case in which $1<M<N$. In this case, $t\in \{1,2,\ldots,K-1\}$. \\
\noindent\textbf{\underline{{Storage Phase:}}}
In the placement phase, each file $W_n$ for $n = 1,\ldots,N$ is split into ${K \choose t}$  non-overlapping sub-files of equal size $F/{K\choose t}$:
\begin{equation}\label{eq:sub}
W_n = (W_{n,\tau} : \tau \subseteq \{1,\ldots,K\}, |\tau| = t).
\end{equation}
\setlength\textfloatsep{0\baselineskip minus 1pt}
\begin{algorithm}[t]
\caption{Secure Centralized Caching Algorithm}
\begin{algorithmic}[1]
\Statex \hspace{5pt}
\Statex \hspace{-15pt}\textbf{Centralized Cache Placement:} for files $W_1,\ldots,W_N$
\State $t ={K(M-1)}/{(N-1)}$
\For{$n \in \{1,2,\ldots,N\}$}
	\State Split file $W_n$ into equal sized fragments $W_{n,\mathcal{T}} : \mathcal{T} \subseteq \{1,2,\ldots,K\}, |\mathcal{T}| = t$ 
\EndFor
\State Generate keys $\K_{\mathcal{T}_k}$ such that $\mathcal{T}_k \subseteq \{1,2,\ldots,K\}, |\mathcal{T}_k| = t+1$
\For{$ k\in \{1,2,\ldots,K\} $}
	\For{$n = 1,2,\ldots,N$}
		\State File $W_{n,\mathcal{T}}$ is place in cache, $Z_k$, of user $k$ if $k \in \mathcal{T}$
		\State Key $\K_{\mathcal{T}_k}$ is placed in cache, $Z_k$, of user $k$ if $k \in \mathcal{T}_k$
	\EndFor
\EndFor
\Statex \hspace{-15pt}\textbf{Coded Delivery:}
\For{$\mathcal{S}$ such that $\mathcal{S}\subseteq\{1,2,\ldots,K\},|\mathcal{S}| = t+1$}
		\State Server sends $\left\{\mathcal{K}_{\mathcal{S}}\oplus_{k\in\mathcal{S}} W_{d_k,\mathcal{S}\setminus\{k\}}\right\}$
\EndFor\vspace{5pt}
\end{algorithmic}\label{alg:0}
\end{algorithm}
For each $n$, the sub-file $W_{n,\tau}$ is placed the cache of user $k$ if $k \in \tau$. Since $|\tau| = t$, for each user $k\in \tau$, there are $t-1$ out of $K-1$ possible users with whom it shares a sub-file of a given file $W_n$. Thus each user caches $N{K-1\choose t-1}$ sub-files. Next we generate a set of keys, each of the size of a sub-file i.e. of size $F/{K\choose t}$: 
\begin{equation}\label{eq:key}
(\mathcal{K}_{\tau_k} : \tau_k \subseteq \{1,\ldots,K\}, |\tau_k| = t+1).
\end{equation}
The key $\mathcal{K}_{\tau_k}$ is placed in the cache of user $k$ if $k\in \tau_k$. The keys are generated such that all the keys are orthogonal to each other and each key is distributed according to $\K_{\tau_k} \sim \textrm{unif} \left\{1,2,\ldots, 2^{{F}/{{K\choose t}}}\right\}$. Again, since $|\tau_k| = t+1$, each user $k\in \tau_k$ shares key $\K_{\tau_k}$ with $t$ out of $K-1$ possible users. Thus there are ${K-1 \choose t}$ keys in the cache of each user. Given each key and sub-file has size $F/{K\choose t}$, number of bits required for storage at each user is:
\begin{align}
&N{N-1\choose t-1}\cdot\frac{F}{{K\choose t}} + {K-1\choose t}\cdot\frac{F}{{K\choose t}} \nonumber\end{align}\begin{align}
=&  \frac{FNt}{K} + F\left(1 - \frac{t}{K}\right) =  F\left(\frac{Nt}{k} + 1 - \frac{t}{K}\right) = FM
\end{align}
which satisfies the memory constraint. \\
\noindent\textbf{\underline{{Delivery Phase:}}}
We now elaborate on the delivery phase. Consider a request vector $(d_1,\ldots,d_k) \in \{1,\ldots,N^K\}$ where user $k$ requests the file $W_{d_k}$. Let $\mathcal{S} \subseteq \{1,\ldots,K\}$ be a subset of $|\mathcal{S}| = t+1$ users. Every $t$ users in $\mathcal{S}$ share a sub-file in their cache which is requested by the $t+1$-th user. Given a user $k\in \mathcal{S}$ and $|\mathcal{S}\setminus\{k\}| = t$, the sub-file $W_{d_k,\mathcal{S}\setminus \{k\}}$ is requested by user $k$ as it is a sub-file of $W_{d_k}$ which is missing at user $k$ since $k\notin \mathcal{S} \setminus \{k\}$.  The file is present in the cache of the $t$ users $s\in \mathcal{S} \setminus \{k\}$. 
 For each such subset $\mathcal{S} \subseteq \{1,\dots,K\}$, the server sends the following transmission: 
$X_{(d_1,\ldots,d_k)} = \left\{\mathcal{K}_{\mathcal{S}}\oplus_{s\in\mathcal{S}} W_{d_s,\mathcal{S}\setminus\{s\}}\right\}$ such that $\left\{\mathcal{S} \subseteq \{1,2,\ldots,K\}, |\mathcal{S}| = t+1 \right\}$.
The number of subsets $\mathcal{S}$ is ${K \choose t+1}$. Thus there are ${K \choose t+1}$ transmissions and an unique key associated with each transmission i.e., there are ${K \choose t+1}$ keys in the system. Each transmission has the size of a subfile and thus the total number of bits sent over the rate-limited link is:
\begin{align}
R_s^C(M)F = & {K\choose t+1}\cdot\frac{F}{{K\choose t}} =\frac{K\left(1 - \frac{M-1}{N-1}\right)}{1 + \frac{K(M-1)}{N-1}}\cdot F\nonumber\end{align}\begin{align}
\Rightarrow R^*_s(M) \leq& R_s^C(M) \triangleq	\frac{K \left(1 - \frac{M-1}{N-1}\right)}{1 + \frac{K(M-1)}{N-1}}.
\end{align}
Next, we show that the delivery phase does not reveal any information to the wiretapper i.e., we show that
\begin{equation}
I\left(X_{(d_1,\ldots,d_k)} ; W_1,\ldots,W_N\right) = 0 
\end{equation}
We have,
\begin{align}
   I\left(X_{(d_1,\ldots,d_K)} ; W_1,\ldots,W_N\right) &\nonumber\\
& \hspace{-120pt}=  H\left(X_{(d_1,\ldots,d_K)}\right) - H\left(X_{(d_1,\ldots,d_K)}|W_1,\ldots,W_N\right)\nonumber\\
& \hspace{-120pt}=  H\left(X_{(d_1,\ldots,d_K)}\right) - H\left(\left\{\mathcal{K}_{\mathcal{S}}\oplus_{s\in\mathcal{S}} W_{d_s,\mathcal{S}\setminus\{s\}}:|\mathcal{S}|=t+1\right\}|W_1,\ldots,W_N\right)\nonumber \\
& \hspace{-120pt}=   H\left(X_{(d_1,\ldots,d_K)}\right) - H\left(\left\{\mathcal{K}_{\mathcal{S}}:|\mathcal{S}|=t+1\right\}|W_1,\ldots,W_N\right)\nonumber \\
& \hspace{-120pt}=   H\left(X_{(d_1,\ldots,d_K)}\right) - H\left(\left\{\mathcal{K}_{\mathcal{S}}:|\mathcal{S}|=t+1\right\}\right),\label{H0} \vspace{-8pt}
\end{align}
where, the last equality follows from the fact that the keys are uniformly distributed and are independent of the files $(W_1,\ldots,W_N)$.  Using the fact that $H(A,B)\leq H(A)+H(B)$, we have:
\begin{align}
 H\left(X_{(d_1,\ldots,d_K)}\right)  &= H\left( \left\{\mathcal{K}_{\mathcal{S}}\oplus_{s\in\mathcal{S}} W_{d_s,\mathcal{S}\setminus\{s\}}:|\mathcal{S}|=t+1\right\}\right)\nonumber\\
												& \leq \sum_{i=1}^{{K \choose t+1}} H\left(\mathcal{K}_{\mathcal{S}_i}\oplus_{s\in\mathcal{S}_i} W_{d_s,\mathcal{S}_i\setminus\{s\}}:|\mathcal{S}_i|=t+1\right)		\nonumber\\
												& \leq \sum_{i=1}^{{K \choose t+1}}\log_2\left(\frac{F}{{K \choose t}}\right) = {K \choose t+1}\log_2\left(\frac{F}{{K \choose t}}\right).\label{H1}
\end{align}
\vspace{-5pt}On the other hand, we have:												
\begin{align}
 H\left(\left\{\mathcal{K}_{\mathcal{S}}:|\mathcal{S}|=t+1\right\}\right) &= \sum_{i=1}^{{K \choose t+1}} H\left(\mathcal{K}_{\mathcal{S}_i}:|\mathcal{S}_i|=t+1\right) \nonumber\\
 & = \sum_{i=1}^{{K \choose t+1}}\log_2\left(\frac{F}{{K \choose t}}\right) = {K \choose t+1}\log_2\left(\frac{F}{{K \choose t}}\right),\label{H2}
\end{align}
where the equality in (\ref{H2}) follows from the fact that the keys $\K_{\mathcal{S}_{i}}$, for all $i$ are mutually independent and distributed as $\textrm{unif} \left\{1,2,\ldots, 2^{{F}/{{K\choose t}}}\right\}$. Substituting (\ref{H1}) and (\ref{H2}) into (\ref{H0}), we have:
\begin{equation}
I\left(X_{(d_1,\ldots,d_K)} ; W_1,\ldots,W_N\right)  \leq 0.\label{H4}
\end{equation}
Using the fact that for any $X,Y$, $I(X;Y)\geq 0$, we have:
\begin{equation}
I\left(X_{(d_1,\ldots,d_K)} ; W_1,\ldots,W_N\right)  = 0,
\end{equation}
which proves that the rate $R^C_s(M)$ is \textit{securely} achievable. This completes the proof of Theorem \ref{th:1}.\qed
%
%
%
%
%
%
%
%
\vspace{-10pt}
\section{Proof of Theorem \ref{th:3}}\label{ssec:th3}
In this section, we prove the information-theoretic lower bound on $R^*_s(M)$ for any $N,K\in\mathbb{N}$. Let $s$ be an integer such that $s\in \{1,\ldots,\min\{N,K\}\}$. Consider the first $s$ caches $Z_1,\ldots,Z_s$. For a request vector $(d_1,d_2,\ldots,d_s,d_{s+1},\ldots,d_K)$ $=$ $ (1,2,\ldots,s,\phi,\ldots,\phi)$, the transmission $X_1 = X_{(d_1,\ldots,d_k)}$, along with the caches $Z_1,\ldots,Z_s$ must be able to decode the files $W_1,\ldots,W_s$. Similarly there for another request $(d_1,d_2,\ldots,d_s,d_{s+1},\ldots,d_K)$ $=$ $(s+1,s+2,\ldots,2s,\phi,\ldots,\phi)$, the transmission $X_2$, which along with caches $Z_1,\ldots,Z_s$, must be able to decode the files $W_{s+1},\ldots,W_{2s}$. Thus considering $\lfloor N/s \rfloor$ different requests, the transmissions from the central server denoted by $X_1,\ldots, X_{\lfloor N/s \rfloor}$, along with the caches $Z_1,\ldots,Z_s$, must be able to decode the files $W_1,\ldots,W_{s\lfloor N/s \rfloor}$. Let 
\begin{align*}
\widetilde{W} &= \left\{W_1,\ldots,W_{s\lfloor N/s \rfloor}\right\}\\
\widetilde{X} &= \left\{X_1,\ldots,X_{\lfloor N/s \rfloor}\right\} \\
\widetilde{X}_{\setminus\{l\}} &= \big\{X_1,\ldots,X_{l-1},X_{l+1},\ldots,X_{\lfloor N/s \rfloor}\big\}\\
\widetilde{Z} &= \left\{Z_1,\ldots,Z_s\right\}.
\end{align*} 
In addition, we also have constraints based on file retrieval and security. The file retrieval constraint is based on the fact that given all possible transmissions and caches, all files can can be retrieved. The security constraint is that a wiretapper should not be able to retrieve any information about the files from any transmission by the server. Using Definition \ref{def1}, we have:\vspace{-2pt}
\begin{align}
&H(\widetilde{W}|\widetilde{X},\widetilde{Z}) \leq \epsilon \label{FR} \\
&I(\widetilde{W};X_l) \leq \epsilon;  ~~~ l = 1,\ldots,\left\lfloor{{N}/{s}}\right\rfloor \label{security}
\end{align}
We present a novel extension to the cut-set bound argument \cite{Cover} to include the security and file retrieval constraints. Consider the information flow consisting of transmissions $X_1,\ldots,X_{\lfloor N/s \rfloor}$ and caches $Z_1,\ldots,Z_s$ for decoding files $W_1,\ldots,W_{s\lfloor N/s \rfloor}$. This flow has minimum capacity $s\lfloor N/s \rfloor$. Thus, we have:
\begin{align}
s\left\lfloor N/s\right\rfloor F & \leq  H(\widetilde{W}) = I(\widetilde{W} ; \widetilde{X}, \widetilde{Z}) + {H(\widetilde{W}|\widetilde{X},\widetilde{Z})}\nonumber\\
															& \myleq{(\ref{FR})} I(\widetilde{W} ; \widetilde{X}, \widetilde{Z}) + \epsilon\nonumber\\
															& =  I\left(\widetilde{W}; \{X_1,\ldots,X_{\lfloor N/s \rfloor}\}, \{Z_1,\ldots,Z_s\}\right) + \epsilon\nonumber \\
															& = {I(\widetilde{W};X_l)} + I\left(\widetilde{W};\widetilde{X}_{\setminus\{l\}},\widetilde{Z}|X_l\right) + \epsilon\nonumber \\
															& \myleq{(\ref{security})} I\left(\widetilde{W};\widetilde{X}_{\setminus\{l\}},\widetilde{Z}|X_l\right) + 2\epsilon \nonumber\\
															&	\leq H\left(\widetilde{X}_{\setminus\{l\}},\widetilde{Z}\right) + 2\epsilon \nonumber\\
															& \leq \sum_{i = 1,i\neq l}^{\lfloor N/s \rfloor} H(X_i) + \sum_{j=1}^{s} H(Z_j) + 2\epsilon\nonumber \\
															& \leq \left(\lfloor N/s \rfloor -1\right) R_s^*(M)F + sMF + 2\epsilon\nonumber \\
\Rightarrow s\left\lfloor N/s \right\rfloor &  \leq \left(\lfloor N/s \rfloor - 1\right)R_s^*(M) + sM + \frac{2\epsilon}{F}. \label{eq:conv}
\end{align}
Solving for $R_s^*$ and optimizing over all possible $s$, we have:
\begin{eqnarray}
R^*_s(M) &\geq& \max_{s\in\{1,\dots,\min\{N,K\}\}} \lim_{\epsilon\rightarrow 0} \frac{s\lfloor{N/s}\rfloor - sM - \frac{2\epsilon}{F}}{\lfloor {N/s}\rfloor - 1}\nonumber\\
				 & = & \max_{s\in\{1,\dots,\min\{N,K\}\}} \left(s - \frac{s(M-1)}{\left(\lfloor \frac{N}{s} \rfloor - 1\right)}\right),
\end{eqnarray}
which concludes the proof of Theorem \ref{th:3}. \qed
%
%
%
%
%
%
%
%
\vspace{-10pt}
\section{Proof of Theorem \ref{th:4}}\label{ssec:th4} 

In this section, we prove that a constant multiplicative gap exists between the securely achievable rate $R^C_s(M)$ given in Theorem \ref{th:1} and the optimal secure rate $R^*_s(M)$, for the regime
\begin{align}\label{regime}
\max\left\{\frac{(K-N)(N-1)}{KN} + 1,1\right\} \leq M \leq N.
\end{align}
We consider two cases for the value of $K$. Firstly, for ${K\leq N}$, we have from Theorem \ref{th:1}:
\begin{equation}\label{eq-min}
R^C_s(M) \leq K\left( 1- \frac{M-1}{N-1}\right)=\min\{ N,K\} \left( 1- \frac{M-1}{N-1}\right).
\end{equation}
For the case of ${K>N}$, (\ref{regime}) reduces to $(K-N)(N-1)/KN + 1\leq M\leq N$. Thus we have:
\begin{align}
						&~~\frac{(K-N)(N-1)}{KN} + 1 \leq M \nonumber\\
\Rightarrow &~~ \frac{1}{N} - \frac{1}{K} \leq \frac{M-1}{N-1} ~~\Rightarrow  K\cdot\frac{1}{1 + K\frac{M-1}{N-1}} \leq N\nonumber\\
\Rightarrow &~~ K\left(1 - \frac{M-1}{N-1}\right)\frac{1}{1 + K\frac{M-1}{N-1}} \leq N\left(1 - \frac{M-1}{N-1}\right)\nonumber\\
\Rightarrow &~~ R^C_s(M) \leq \min\{N,K\}\left(1 - \frac{M-1}{N-1}\right). \label{eq:minnkub}
\end{align}
\noindent To prove the constant gap result, we focus on two cases:

\noindent \underline{\textbf{Case $1$:}} $\min\{N,K\}\leq 17$

Setting $s=1$ in Theorem \ref{th:3} gives the following lower bound on the optimal secure rate:
\begin{equation} \label{eq:minnklb}
R^*_s(M)\geq  \left( 1- \frac{M-1}{N-1}\right).
\end{equation}
Hence from (\ref{eq:minnkub}) and (\ref{eq:minnklb}), we have
\begin{equation}
\frac{R^C_s(M)}{R^*_s(M)} \leq \min\{N,K\} \leq 17.
\end{equation}
\noindent \underline{\textbf{Case $2$:}} $\min\{N,K\}\geq18$

For this case, the rate in Theorem \ref{th:1} has 3 distinct regimes:  
\begin{itemize}
	\item \underline{\textbf{Regime $1$:}~~}\\
	 $~~\max\left\{\frac{(K-N)(N-1)}{KN},0 \right\} \leq M - 1\leq 1.2\max\left(1, \frac{N-1}{K}\right)$
	\item \underline{\textbf{Regime $2$:}} $1.2\max\left(1, \frac{N-1}{K}\right) < M - 1 \leq 0.0628 (N-1)$ 
	\item \underline{\textbf{Regime $3$:}} $0.0628(N-1) < M-1\leq N-1$
\end{itemize}

\noindent We consider each of the three regimes separately.
\vspace{5pt}

\noindent\underline{\textbf{Regime $1$:}} $\max\left\{\frac{(K-N)(N-1)}{KN},0 \right\}  \leq M - 1 \leq 1.2\max\left(1,\frac{N-1}{K}\right)$

\noindent By Theorem \ref{th:1}, we have: 
\begin{equation}\label{r1}
R^C_s(M)\leq R^C_s(1) \leq \min\{N,K\}.
\end{equation}
By Theorem \ref{th:3} and using the fact that $\lfloor N/s \rfloor \geq N/s - 1$, we have:
\begin{equation}
R^*_s(M) \geq s- \frac{s^2(M-1)}{N - 2s}.
\end{equation}
Setting $s = \lfloor 0.1586 \min\{N,K\}\rfloor \in \{1,\ldots,\min\{N,K\}\}$ we get, for $M-1\leq 1.2 \max \left(1, \frac{N-1}{K}\right)$: 
\begin{align}
\hspace{-10pt}R^*_s(M) & \geq R^*_s\left(1.2\max \left(1, \frac{N-1}{K}\right) + 1\right)\nonumber\\
				 & \geq 0.1586\min\{N,K\} - 1 -\frac{(0.1586\min\{N,K\})^2\cdot 1.2\max\left(1,\frac{N-1}{K}\right)}{N - 2\cdot 0.1586\min\{N,K\}} \nonumber\end{align}\begin{align}
				 & \geq \min\{N,K\}\Bigg\{ 0.1586 - \frac{1}{\min\{N,K\}} - \frac{(0.1586)^2\cdot 1.2}{1 - 2\cdot(0.1586)\min\{1,K/N\}}\Bigg\}\nonumber\\ 
				 & \geq \min\{N,K\}\left\{0.1586 - \frac{1}{18} - \frac{1.2\cdot(0.1586)^2}{1-2\cdot0.1586} \right\}\nonumber\\
				 & \geq \frac{1}{17} \min\{N,K\}.\label{r2}
\end{align}
Combining (\ref{r1}) and (\ref{r2}), we have:
\begin{equation}
\frac{R^C_s(M)}{R^*_s(M)} \leq 17.
\end{equation}

\noindent\underline{\textbf{Regime $2$:}} $1.2\max\left(1, \frac{N-1}{K}\right) < M - 1 \leq 0.0628 (N-1)$\\ 

Let $\bar{M}$ be the largest multiple of $\frac{N-1}{K}$ less than equal to $M$ such that
\begin{equation}
0\leq M - \frac{N-1}{K} \leq \bar{M} \leq M.
\end{equation}
Choosing $\bar{M} = M - (N-1)/K$, and using the fact that $R^C_s(M)$ is monotonically decreasing in $M$, we have:
\begin{align}
R^C_s(M)&\leq R^C_s(\bar{M})\nonumber\\
&\leq K\cdot{\left\{  1 - \frac{M-1}{N-1} + \frac{1}{K} \right\}}\cdot\frac{1}{1 + \frac{K(M-1)}{N-1} - 1} \leq \left( \frac{N-1}{M-1}\right),\label{r3}
\end{align}
where we have used $\frac{M-1}{N-1}> \frac{1}{K}$ in the last inequality. Now setting $s = \lfloor 0.1530 \frac{N-1}{M-1}\rfloor \in \{1,\ldots,\min\{N,K\}\}$ in Theorem \ref{th:3}, we have: 
\begin{align}
R^*_s(M) & \geq 0.1530\frac{N-1}{M-1} - 1 - \frac{0.1530^2\cdot{\frac{N-1}{M-1}}^2\cdot(M-1) }{N - 2\cdot0.1530\cdot\frac{N-1}{M-1}}\nonumber\\
				 & \geq \frac{N-1}{M-1}\left\{ 0.1530 - 0.0628 - \frac{0.1530^2}{1 - \frac{2\cdot0.1530}{1.2}}\right\}\nonumber\\
				 & \geq \frac{1}{17}\left(\frac{N-1}{M-1}\right). \label{r4} 
\end{align}
Combining (\ref{r3}) and (\ref{r4}), we get:
\begin{equation}
\frac{R^C_s(M)}{R^*_s(M)} \leq 17.
\end{equation}

\noindent\underline{\textbf{Regime $3$:}}  $0.0628(N-1) < M-1\leq N-1$\\

Let $\bar{M}- 1$ be a multiple of $(N-1)/K$ less than equal to $0.0628 (N-1)$, such that
\begin{equation}\label{reg3}
0\leq 0.0628(N-1) - \frac{N-1}{K} \leq \bar{M} - 1 \leq 0.0628(N-1).
\end{equation}
Then using Theorem \ref{th:1} and the fact that $\bar{M}\leq M$, we have: 
\begin{align}\label{60}
&{R^C_s({M})}\cdot\frac{1}{1 - \frac{M-1}{N-1}} \leq {R^C_s(\bar{M})}\cdot\frac{1}{1 - \frac{\bar{M}-1}{N-1}} \nonumber\\
\Rightarrow&R^C_s(M)  \leq {R^C_s(\bar{M})}\cdot\frac{1}{1 - \frac{\bar{M}-1}{N-1}}\cdot \left(1 - \frac{M-1}{N-1}\right)\nonumber\\
			 & \hspace{33pt}\leq {R^C_s(\bar{M})}\cdot\frac{1}{1 - 0.0628}\cdot \left(1 - \frac{M-1}{N-1}\right).
\end{align}
Now by Theorem \ref{th:1} and using (\ref{reg3}), we have:
\begin{align}\label{61}
R^C_s(\bar{M}) &\leq \frac{1}{\frac{\bar{M}-1}{N-1} + \frac{1}{K}} \leq \frac{1}{0.0628 - \frac{1}{K} + \frac{1}{K}} = \frac{1}{0.0628}. 
\end{align}
Thus we have, from (\ref{60}) and (\ref{61}):
\begin{equation}\label{r5}
R^C_s(M) \leq \frac{1}{0.0628(1-0.0628)}\left(1 - \frac{M-1}{N-1}\right).
\end{equation}
Setting $s=1$ in Theorem \ref{th:3}, we have the following lower bound:
\begin{equation}\label{r6}
R^*_s(M) \geq \left(1 - \frac{M-1}{N-1}\right).
\end{equation}
Thus combining (\ref{r5}) and (\ref{r6}), we get:
\begin{equation}
\frac{R^C_s(M)}{R^*_s(M)} \leq \frac{1}{0.0628(1-0.0628)} \leq 17.
\end{equation}
Thus we have proved that for any $N,K \in \mathbb{N}$ and all $\frac{(K-N)(N-1)}{KN} + 1 \leq M\leq N$, there is a constant multiplicative gap of $17$ between the achievable rate and the information theoretic optimal. This concludes the proof of Theorem \ref{th:4}. 
\begin{remark}
For $K\leq N$ the gap is bounded for the entire feasible regime of $1\leq M \leq N$. However, for $K>N$, the gap is unbounded in the regime:
\begin{align*}
1\leq M < \frac{(K-N)(N-1)}{KN} + 1, 
\end{align*}
and scales with the number of users $K$. However, $\frac{(K-N)(N-1)}{KN}  \leq 1$ for any $K>N$ and thus the regime is a fraction of the value of $M$ and is in general negligible when $N$ is large. Also, the regime is always below the values of $M$ for which the data memory dominates key memory i.e., $M> 2N/(N+1) \geq 1$, thereby making it a regime of lesser practical interest. \qed
\end{remark}
%
%
%
%
%
%
%
%
\vspace{-8pt}
\section{Proof of Theorem \ref{th:5}}\label{ssec:th5}

The decentralized algorithm which achieves the rate in Theorem \ref{th:5} is given in Algorithm \ref{alg:1}. 
\begin{algorithm}[h] \vspace{10pt}
\caption{Secure Decentralized Caching Algorithm}
\begin{algorithmic}[1]
\Statex \hspace{10pt}
\Statex \hspace{-15pt}\textbf{Decentralized Cache Placement:}
\For{$k \in \{1,\ldots, K\}, n\in \{1,\ldots,N\}$}
	\State User $k$ randomly caches $\frac{M-1}{N-1}F$ bits of file $n$.
\EndFor
\Statex \hspace{-15pt}\textbf{Delivery Procedure} for request $(d_1,\ldots,d_K)$
\Statex \hspace{-15pt}\textit{Centralized Key Placement:}
\Statex \hspace{-15pt}\textit{Central server maps the cache contents to fragments in the } 
\Statex \hspace{-15pt}\textit{files $W_1,\ldots,W_N$ and generates keys as follows-}
\For{$i = 0,1,2,\ldots,K$}
	\For{$n=1,2,\ldots,N$}
		\State 	$W_n = \{W_{n,\mathcal{T}}\},~~ \mathcal{T} \subseteq \{1,\ldots,K\} : |\mathcal{T}| = i $ such that $W_{n,\mathcal{T}}$ is cached at user $k$, if $k\in \{\mathcal{T}\}$
	\EndFor
\EndFor
\For{$s = 1,2,\ldots,K$}
	\For{$\mathcal{S} \subseteq \{1,\ldots,K\} : |\mathcal{S}| = s$}
		\State Key $\K_{\mathcal{S}}$ is generated
		\State $\K_{\mathcal{S}}$ is placed in cache of user $k$ if $k\in \{\mathcal{S}\}$
	\EndFor
\EndFor

\Statex \hspace{-15pt}\textit{Coded Secure Delivery:}
\For{$s = K,K-1,\ldots,1$}
	\For{ $\mathcal{S} \subseteq \{1,\ldots, K\} : |\mathcal{S}| = s$}
		\State Server sends $\left\{\mathcal{K}_{\mathcal{S}}\oplus_{k\in\mathcal{S}} W_{d_k,\mathcal{S}\setminus\{k\}}\right\}$
	\EndFor
\EndFor

\Statex \hspace{-15pt}\textbf{Conventional Delivery Procedure} for request $(d_1,\ldots,d_K)$
\State Server places individual keys of size $(1 - \frac{M-1}{N-1})F$ bits at each user's cache
\For{$n\in \{0,\ldots,N\}$}
	\State Server sends enough random linear combinations of bits in file $n$ XOR-ed with individual keys for the all users requesting it
\EndFor \vspace{3pt}
\end{algorithmic}\label{alg:1}
\end{algorithm}

Given $N$ files and $K$ users, each with a cache size of $MF$ bits, we first show that the memory constraint $M \in \frac{N-1}{N}t + 1$ for $t \in (0,N]$ is valid. We then evaluate the rate of Algorithm \ref{alg:1} and show that the multicast delivery is information theoretically secure. 
 
Considering the proposed decentralized scheme in Algorithm \ref{alg:1}, each user is allowed to cache any random subset of $\frac{M-1}{N-1}F$ bits of any file $W_n$. Since the choice of these subsets is uniform, given a particular bit in file $W_n$, the probability of the bit being cached at a given user is: 
\begin{equation}\label{q}
q \triangleq \frac{M-1}{N-1} \in (0,1]. 
\end{equation}
Considering a fixed subset of $s$ out of $K$ users, the probability that this bit is cached exactly at these $s$ users and not cached at the remaining $(K-s)$ users is $q^s(1-q)^{K-s}$. The expected number of bits of $W_n$ that are cached at exactly those $s$ users is given by:
\begin{align}\label{preveq}
E\left[\textrm{\# of bits of} ~W_n ~\textrm{ at $s$ users}\right] = Fq^s(1-q)^{K-s}. 
\end{align}
The actual realization of the random number of bits of a file $W_n$ cached at $s$ users is within the range: 
\begin{equation}\label{frag}
Fq^{s}(1-q)^{K-s} \pm o(F).
\end{equation} 
For ease of exposition, we consider all the fragments of files shared by $s$ users have the same size. Hence the factor $o(F)$ can be ignored for large enough $F$. 

\subsection*{Memory Constraint}
Next, the server maps the contents of the users' caches to non-overlapping fragments in files such that each fragment reflects which users have cached the bits contained in the fragment. Referring to Algorithm \ref{alg:1}, Line 4, the variable $i$ signifies the number of users which share a given file fragment. For $i=0$, the file fragments are $W_{n,\phi}$ which is not stored at any user. When $i=1$, the file fragments are $W_{n,k}$ for $k=1,\ldots,K$ which are stored only at one user and hence shared by none. In general for any $i$, the fragments $W_{n,\mathcal{S}}$ such that $|\mathcal{S}| = i$ are stored at $i$ users and shared by any given user with $i-1$ other users. Thus, for a given a user $k$, the number of fragments it shares with $i-1$ out of the remaining $K-1$ users for each $i$ is given by ${K-1 \choose i-1}$. From (\ref{preveq}), we have the size of fragments which are stored at exactly $i$ users is $Fq^i(1-q)^{K-i}$. Thus, the total memory at each user for storing data is given by:
\begin{align}
M_DF &= N\cdot\sum_{i=1}^{K}{K-1 \choose i-1}Fq^i(1-q)^{K-i} \nonumber\end{align}\begin{align}
M_D~~&= Nq \sum_{i-1=0}^{K-1}{K-1 \choose i-1}q^{i-1}(1-q)^{(K-1)-(i-1)} \nonumber\\
		&= Nq = N\frac{M-1}{N-1}. \label{md}
\end{align}
Next, we describe the centralized key placement. For each sub-set $\mathcal{S} \subseteq \{1,\ldots,K\}$ of size $s$, i.e., $|\mathcal{S}| = s$, where $s=1,2,\ldots,K$, a key $\K_{\mathcal{S}}$ is generated as follows:

\begin{align}
\K_{\mathcal{S}}\sim \textrm{unif}\left\{1, 2, \ldots, 2^{Fq^{s-1}(1-q)^{K-s+1}}\right\}.\label{KeyGenDecen} 
\end{align}
Subsequently, the key $\K_{\mathcal{S}}$ is placed in the cache of user $k$ if $k\in \mathcal{S}$. The centralized key generation and placement phase is inherently related to the delivery phase of the decentralized algorithm since the size of a key is related to the size of file fragment which is encoded with the key during coded delivery. Consider the coded delivery phase in Algorithm \ref{alg:1}, Line $15-19$. Given a request $(d_1,\ldots,d_K)$, the composite transmission $X_{(d_1,\ldots,d_K)}$ is sent by the server. The composite transmission can be written as:
\begin{equation}\label{Xd}
X_{(d_1,\ldots,d_K)} = \left\{X^s_{(d_1,\ldots,d_K)}\right\}_{s=1}^{K},
\end{equation}
where $X^s_{(d_1,\ldots,d_K)}$ consists of ${K \choose s}$ transmissions, one for each possible sub-set $\mathcal{S}$ of size $s$ i.e., 
\begin{align}\label{xs}
X^s_{(d_1,\ldots,d_K)} = \left\{\K_{\mathcal{S}}\oplus_{k\in\mathcal{S}} W_{d_k,\mathcal{S}\setminus\{k\}}:  |\mathcal{S}| = s\right\}. 
\end{align}
$W_{d_k,\mathcal{S}\setminus\{k\}}$ denotes the part of the file $W_{d_k}$ requested by user $k$ which is present in the caches all the users in set $\mathcal{S}$ except in the cache of user $k$. The key $\K_{\mathcal{S}}$ is associated with the transmission $\oplus_{k\in\mathcal{S}} W_{d_k,\mathcal{S}\setminus\{k\}}$. Furthermore, from the design of the key placement, the key $\K_{\mathcal{S}}$ is available in the cache of all the $s$ users in the sub-set $\mathcal{S}$.
Since $|\mathcal{S}\setminus \{k\}| = s - 1$, from (\ref{preveq}) we have, the expected size of the fragment $W_{d_k,\mathcal{S}\setminus\{k\}}$ is given by $Fq^{s-1}(1-q)^{K-s+1}$. For a fixed value of $s$, the size of each transmission in $X^s_{(d_1,\ldots,d_K)}$ is given by: 
\begin{equation}\label{fsize}
\max_{k\in \mathcal{S}} |W_{d_k,\mathcal{S}\setminus\{k\}}| = Fq^{s-1}(1-q)^{K-s+1}.
\end{equation}
Thus, each key $\K_{\mathcal{S}}$ must be chosen with the size:
\begin{equation}
|\K_{\mathcal{S}}| = \max_{k\in \mathcal{S}} |W_{d_k,\mathcal{S}\setminus\{k\}}| =  Fq^{s-1}(1-q)^{K-s+1},
\end{equation}
which is precisely how each key is generated according to (\ref{KeyGenDecen}). 
Now, for a given value of $s$, a user $k$ needs file fragments contained in $\mathcal{S}\setminus\{k\}$ i.e., $s-1$ other users in the set $\mathcal{S}$. This set of $s-1$ users need to be chosen out of the remaining $K-1$ users. Thus for each $s$, there are ${K-1 \choose s-1}$ keys associated with each user. Thus the total number of keys at each user is given by $\sum_{s=1}^{K} {K-1 \choose s-1} = 2^{K - 1}$. The total memory occupied by keys at each users' cache is given by:
\begin{align}
M_KF  &=  \sum_{s=1}^{K} {K-1 \choose s-1} Fq^{s-1}(1-q)^{K-s+1} \nonumber\\\label{mk}
M_K~~ &=  (1-q)\sum_{s=1}^{K} {K-1 \choose s-1} Fq^{s-1}(1-q)^{(K-1)-(s-1)} \nonumber\\
		  &=  (1-q) = 1 - \frac{M-1}{N-1}.
\end{align}
From (\ref{mk}) and (\ref{md}), we have:  
\begin{align}
 M_D + M_K = N\frac{M-1}{N-1} + 1 - \frac{M-1}{N-1} = M,  
\end{align}
which proves the memory constraint. Putting $M_D = t$, the memory break up can be parametrized as:
\begin{align}
M = t + (1 - \frac{t}{N}) = \frac{N-1}{N}t + 1. 
\end{align}
Now, when $t=0$, $M = 1$, which is the condition for storing just keys in caches and sending entire files over the shared link. On the other hand, when $t=N$, $M=N$ i.e., the entire files are stored in the caches and there is no need for a transmission. Thus $t\in (0,N]$ is the region of interest. Hence $M\in \frac{N-1}{N}\cdot(0,N] + 1$ is valid. Note that the constraint on $M$ is due to the centralized key placement and is thus the cost for security. 
\begin{remark}
Considering the range for file fragment size in (\ref{frag}), if we consider that the fragments are not indeed of equal size, then in turn the key size is also within the range $M_K \pm o(F)$. If this is the case, then the cache memory constraint will be within the range $M \pm o(F)$. Since $o(F)$ can generally be ignored in comparison to $M$, the cache memory constraint is satisfied on an average. \hfill $\Diamond$ 
\end{remark}
\vspace{-10pt}
\subsection*{Calculation of $R^D_s(M)$}
\subsection{Analysis of Conventional Secure Scheme}
In conventional secure delivery scheme, for $N\leq K$, the worst case request corresponds to at least one user requesting every file. Considering all users request file $W_n$, they all have $F(M-1)/(N-1)$ of its bits already in their cache. Thus at most $F\left(1-\frac{M-1}{N-1}\right) + o(F)$ random linear combinations need to be sent to the users requesting the file $n$. For ease of exposition, $o(F)$ can be ignored. In the conventional scheme, each user $k$ stores an unique key $\K_{k}$ of size $\left(1  - \frac{M-1}{N-1}\right)F$ bits which is XOR-ed with the data before transmission. {Although there are $N$ files, each users' request needs to be secured with a key. Thus, in contrast to the non-secure case in \cite{Maddah-Ali-decentralized}, the unicast delivery is done for $K$ users and the normalized delivery rate is $K\left(1 - \frac{M-1}{N-1}\right).$}

If $N>K$, then at most $K$ different files can be requested. The transmission thus has a normalized rate of $K\left(1- \frac{M-1}{N-1}\right)$. Thus, for all $N$ and $M\in(1,N]$, the conventional scheme has a normalized rate of: 
\begin{align}\label{conv}
R^{\text{conv}}_s(M) & = K\left(1 - \frac{M-1}{N-1}\right) 
\end{align}\vspace{-15pt}
\subsection{Analysis of the proposed scheme}
Considering the secure delivery procedure for the coded caching scheme in Algorithm \ref{alg:1}, we can see that there are ${K \choose s}$ subsets $\mathcal{S}$ of cardinality $s$. Thus there are ${K \choose s}$ transmissions for each $s = K,K-1,\ldots,1$. 
Now, for the coded secure transmission, the unique key $\mathcal{K}_{\mathcal{S}}$ is associated with each subset $\mathcal{S}$. The total number of unique keys in the system is given by $\sum_{s=1}^{K} {K \choose s} = 2^K - 1$.

Now, considering the fragment size of $W_{d_k,\mathcal{S}\setminus\{k\}}$ in (\ref{fsize}) and the transmission $X^s_{(d_1,\ldots,d_K)}$ in (\ref{xs}), for each value of $s$, the size of each transmission is given by:
\begin{equation}
|X^s_{(d_1,\ldots,d_K)}| = {K \choose s}Fq^{s-1}(1-q)^{K-s+1}.
\end{equation}
Summing over all values of $s$, the rate $R^{\text{dec}}_s(M)$, of the composite transmission $X_{(d_1,\ldots,d_K)}$ is:
\begin{align}
\hspace{-10pt}R^{\text{dec}}_s(M)F 		 &= \sum_{s=1}^{K}  {K \choose s}Fq^{s-1}(1-q)^{K-s+1} \nonumber \\
\hspace{-10pt}R^{\text{dec}}_s(M)~~ &=  \frac{1-q}{q}\cdot \sum_{s=1}^{K}{K \choose s} q^{s}(1-q)^{K-s} \nonumber\\
					& =  \frac{1-q}{q}\cdot\left(1- (1-q)^K\right) \nonumber\\
					& \myeq{(\ref{q})}  \frac{1 - \frac{M-1}{N-1}}{\frac{M-1}{N-1}}\cdot\left(1-\left(1- \frac{M-1}{N-1}\right)^K\right)\nonumber \end{align}\begin{align}
					& =  K\left( 1 - \frac{M-1}{N-1}\right)\cdot\frac{N-1}{K(M-1)}\cdot\left(1-\left(1- \frac{M-1}{N-1}\right)^K\right). \label{pr:th5}
\end{align}
{The server can use either the proposed scheme or the conventional secure scheme, whichever uses the minimal rate. Thus combining (\ref{conv}) and (\ref{pr:th5}), Algorithm \ref{alg:1} achieves a rate of:
\begin{align}\label{rds}
R_s^D(M) &= \min\left\{ R^{\text{conv}}_s(M),R^{\text{dec}}_s(M)\right\} \nonumber\\
&= K\left( 1 - \frac{M-1}{N-1}\right) \cdot \nonumber\\
&~~~\min\left\{  \frac{N-1}{K(M-1)}\cdot\left(1-\left(1- \frac{M-1}{N-1}\right)^K\right), 1\right\},
\end{align}
which is the result (\ref{eq:th5}) presented in Theorem \ref{th:5}.}

\subsection*{Proof of Secure Achievability}
Next, we show that the delivery phase does not reveal any information to the wiretapper i.e., we show that:
\begin{equation}
I\left(X_{(d_1,\ldots,d_K)} ; W_1,\ldots,W_N\right) = 0 
\end{equation}

In the decentralized scheme, the central server transmits $X_{(d_1,\ldots,d_K)}$ to satisfy the requests $(d_1,\ldots,d_k)$ of the $K$ users. The composite transmission $X_{(d_1,\ldots,d_K)}$, given in (\ref{Xd}), consists of ${K\choose s}$ transmissions for each $s = K,K-1,\ldots,1$. We have:
\begin{align}
& I\left(X_{(d_1,\ldots,d_K)} ; W_1,\ldots,W_N\right)  \nonumber\\
&=  H\left(X_{(d_1,\ldots,d_K)}\right) - H\left(X_{(d_1,\ldots,d_K)}|W_1,\ldots,W_N\right)\nonumber\\
&=  H\left(X_{(d_1,\ldots,d_K)}\right) - H\left(\left\{X^s_{(d_1,\ldots,d_K)}\right\}_{s=1}^{K}|W_1,\ldots,W_N\right)\nonumber\\
&=  H\left(X_{(d_1,\ldots,d_K)}\right) - H\left(\left\{\left\{\mathcal{K}_{\mathcal{S}}\oplus_{k\in\mathcal{S}} W_{d_k,\mathcal{S}\setminus\{k\}}:{|\mathcal{S}| = s}\right\}\right\}_{s=1}^{K}|W_1,\ldots,W_N\right)\nonumber\\
&=  H\left(X_{(d_1,\ldots,d_K)}\right) - H\left(\left\{\left\{\mathcal{K}_{\mathcal{S}}:{|\mathcal{S}| = s}\right\}\right\}_{s=1}^K|W_1,\ldots,W_N\right)\nonumber\\
&=  H\left(X_{(d_1,\ldots,d_K)}\right) - H\left( \left\{\left\{\mathcal{K}_{\mathcal{S}} :{|\mathcal{S}| = s}\right\}\right\}_{s=1}^{K}\right), \label{H5}
\end{align}
where, the last equality follows from the fact that the keys are uniformly distributed and are independent of the files $W_1,\ldots,W_N$. Using the fact that $H(A,B)\leq H(A)+H(B)$, we have:
\begin{align}
&H\left(X_{(d_1,\ldots,d_K)}\right) \nonumber\\
& = H\left(\left\{X^s_{(d_1,\ldots,d_K)}\right\}_{s=1}^{K} \right)  \leq \sum_{s=1}^{K}  H\left(X^s_{(d_1,\ldots,d_K)}\right) \nonumber \end{align}\begin{align}
												& \leq \sum_{s=1}^{K}\sum_{i=1}^{{K \choose s}} H\left(\mathcal{K}_{\mathcal{S}_i}\oplus_{k\in\mathcal{S}_i} W_{d_k,\mathcal{S}_i\setminus\{k\}}:{|\mathcal{S}_i| = s}\right) \nonumber\\
												& \leq \sum_{s=1}^{K}\sum_{i=1}^{{K \choose s}}\log_2\left(Fq^{s-1}(1-q)^{K-s+1}\right) \nonumber\\
												& = \sum_{s=1}^{K}{K \choose s}\log_2\left(Fq^{s-1}(1-q)^{K-s+1}\right). \label{H6}
												\end{align}												
On the other hand, we have:											
\begin{align}
&H \left( \left\{\left\{\mathcal{K}_{\mathcal{S}}:{|\mathcal{S}| = s}\right\}\right\}_{s=1}^{K}\right) \nonumber\\
&= \sum_{s=1}^{K} H\left(\left\{\mathcal{K}_{\mathcal{S}}:{|\mathcal{S}| = s}\right\}\right) = \sum_{s=1}^{K}\sum_{i=1}^{{K \choose s}} H\left(\mathcal{K}_{\mathcal{S}_i}:{|\mathcal{S}_i| = s}\right) \nonumber\\
&= \sum_{s=1}^{K}\sum_{i=1}^{{K \choose s}}\log_2\left(Fq^{s-1}(1-q)^{K-s+1}\right) \nonumber\\
&= \sum_{s=1}^{K}{K \choose s}\log_2\left(Fq^{s-1}(1-q)^{K-s+1}\right),\label{H7}
\end{align}
where the equality in (\ref{H7}) follows from the fact that the keys are orthogonal to each other and they are uniformly distributed as in (\ref{KeyGenDecen}). Substituting (\ref{H6}) and (\ref{H7}) into (\ref{H5}), we have:
\begin{equation}
I\left(X_{(d_1,\ldots,d_K)} ; W_1,\ldots,W_N\right)  \leq 0\label{H8}
\end{equation}
Using the fact that for any $X,Y$, $I(X;Y)\geq 0$, we have:
\begin{equation}
I\left(X_{(d_1,\ldots,d_K)} ; W_1,\ldots,W_N\right)  = 0
\end{equation}
which proves that the rate $R^D_s(M)$ is \textit{securely} achievable. This completes the proof of Theorem \ref{th:5}.\qed\\ 

\section{Proof of Theorem \ref{th:6}}\label{ssec:th6}
The proof for Theorem \ref{th:6} is similar to the proof of Theorem \ref{th:4} in Appendix \ref{ssec:th4}. We prove that a constant multiplicative gap exists between the achievable decentralized secure rate in Theorem \ref{th:5} and the information theoretic optimal for the regime: 
\begin{align}
\frac{N-1}{N} + 1\leq M \leq N
\end{align}

\noindent For the case of ${K<N}$, from Theorem \ref{th:5}, we have, for $1<M\leq N$,
\begin{align}\label{eqmindec}
R_s^D(M) \leq K\left( 1 - \frac{M-1}{N-1}\right) = \min\{N,K\}\left( 1 - \frac{M-1}{N-1}\right).
\end{align}

\noindent Again in the case of ${K>N}$, we have
\begin{align} \label{cond1}
& M \geq \frac{N-1}{N} + 1 \Rightarrow \frac{N-1}{M-1} < N
\end{align}
\noindent Now, setting $r = 1 - \frac{M-1}{N-1}$ and substituting in (\ref{cond1}), we have:
\begin{align}
\frac{1}{1-r} < N
\end{align}
\noindent Since $0\leq r < 1$, we have
\begin{align}\label{geom}
\frac{1}{1-r}  \approx \sum_{i = 0}^{K-1} r^i \leq N,
\end{align}
which becomes tighter as $K \rightarrow \infty $. Noting that (\ref{geom}) is a geometric series, we get:
\begin{align}
\sum_{i = 0}^{K-1} r^i \leq N ~~\Rightarrow~~  \frac{1 - r^K}{1-r} \leq N 
\end{align}
Substituting the value of $r$, we have:
\begin{align}\label{rdub}
&\frac{N-1}{M-1}\left(1-\left(1- \frac{M-1}{N-1}\right)^K\right) \leq N \nonumber\\
\Rightarrow ~~~& R_s^D(M) \leq \min\{N,K\}\left( 1 - \frac{M-1}{N-1}\right)
\end{align}
\noindent Thus in general, $R_s^D(M) \leq \min\{N,K\}\left( 1 - \frac{M-1}{N-1}\right)$ for the regime:
\begin{align}
 \frac{N-1}{N} + 1\leq M \leq N.
\end{align}


%

\noindent Next, we consider two cases: $\min\{N,K\}\leq 17$ and $\min\{N,K\}\geq 18$.\\
\underline{\textbf{Case $1$:}} $\min\{N,K\}\leq 17$

From (\ref{rdub}), we have:
\begin{align}
R^D_s(M) \leq \min\{ N,K\} \left( 1- \frac{M-1}{N-1}\right).
\end{align}
Also, setting $s=1$ in Theorem \ref{th:3} gives:
\begin{equation}
R^*_s(M)\geq  \left( 1- \frac{M-1}{N-1}\right).
\end{equation}
Thus we have: 
\begin{equation}
\frac{R^D_s(M)}{R^*_s(M)} \leq \min\{N,K\} \leq 17.
\end{equation}

\noindent For $\min\{N,K\}\geq18$,  we consider 3 distinct regimes:  \vspace{5pt}
\begin{equation*}
\begin{array}{l}
\hspace{-250pt}\underline{\textbf{Regime 1:}}~~~ \frac{N-1}{N} + 1 \leq M - 1 \leq 1.2\max\left(1, \frac{N-1}{K}\right)\\
\hspace{-250pt}\underline{\textbf{Regime 2:}}~~~ 1.2\max\left(1, \frac{N-1}{K}\right) < M - 1 \leq \frac{(N-1)}{17} \\
\hspace{-250pt}\underline{\textbf{Regime 3:}}~~~ \frac{(N-1)}{17} < M-1 \leq N-1\\
\end{array}
\end{equation*}

\noindent We consider each of the three regimes separately.\\

\noindent\underline{\textbf{Regime 1:}} $\frac{N-1}{N} + 1 \leq M - 1 \leq 1.2\max\left(1, \frac{N-1}{K}\right)$\\
By (\ref{rdub}), we have: 
\begin{equation}\label{r1d}
R^D_s(M)\leq R^D_s(1) \leq \min\{N,K\}.
\end{equation}
By Theorem \ref{th:3} and using the fact that $\lfloor N/s \rfloor \geq N/s - 1$, we have: 
\begin{equation}
R^*_s(M) \geq s- \frac{s^2(M-1)}{N - 2s}.
\end{equation}
Setting $s = \lfloor 0.1586\min\{N,K\}\rfloor$ we get, for $M-1\leq 1.2 \max \left(1, \frac{N-1}{K}\right)$:
\begin{align}
R^*_s(M) & \geq R^*_s\left(1.2\max \left(1, \frac{N-1}{K}\right) + 1\right)\nonumber\\
				 & \geq 0.1586\min\{N,K\} - 1 -\frac{(0.1586\min\{N,K\})^2\cdot 1.2\max\left(1,\frac{N-1}{K}\right)}{N - 2\cdot 0.1586\min\{N,K\}} \nonumber\\
				 & \geq \min\{N,K\}\left\{ 0.1586 - \frac{1}{\min\{N,K\}} -\frac{(0.1586)^2\cdot 1.2}{1 - 2\cdot(0.1586)\min\{1,K/N\}}\right\}\nonumber\\
				 & \geq \min\{N,K\}\left\{0.1586 - \frac{1}{18} - \frac{1.2\cdot(0.1586)^2}{1-2\cdot0.1586} \right\} \nonumber\\
				 & \geq \frac{1}{17} \min\{N,K\}.\label{r2d}
\end{align}
Combining (\ref{r1d}) and (\ref{r2d}), we get:
\begin{equation}
\frac{R_s^D(M)}{R^*_s(M)} \leq 17.
\end{equation}

\noindent\underline{\textbf{Regime 2:}} $1.2\max\left(1, \frac{N-1}{K}\right) < M - 1 \leq \frac{(N-1)}{17}$\\

Using (\ref{rdub}), we have:
\begin{align}\label{r3d}
R^D_s(M) \leq \frac{N-1}{M-1} - 1 \leq \frac{N-1}{M-1}.
\end{align}

Now setting $s = \lfloor 0.1460 \frac{N-1}{M-1}\rfloor $ in Theorem \ref{th:3}, we have:
\begin{align}
R^*_s(M) & \geq 0.1460\frac{N-1}{M-1} - 1 - \frac{0.1460^2\cdot{\frac{N-1}{M-1}}^2\cdot(M-1) }{N - 2\cdot0.1460\cdot\frac{N-1}{M-1}}\nonumber\\
				 & \geq \frac{N-1}{M-1}\left\{ 0.0.1460 - \frac{1}{17} - \frac{0.1460^2}{1 - \frac{2\cdot 0.1460}{1.2}}\right\}\nonumber\\
				 & \geq \frac{1}{17}\left(\frac{N-1}{M-1}\right). \label{r4d} 
\end{align}
Combining (\ref{r3d}) and (\ref{r4d}), we get:
\begin{equation}
\frac{R^D_s(M)}{R^*_s(M)} \leq 17.
\end{equation}

\noindent\underline{\textbf{Regime 3:}}  $\frac{(N-1)}{17} < M-1 \leq N-1$\\

From (\ref{rdub}), we have:
\begin{align}\label{r5d}
R^D_s(M) \leq \frac{N-1}{M-1} - 1.
\end{align}

Setting $s=1$ in Theorem \ref{th:3}, we have again: 
\begin{equation}\label{r6d}
R^*_s(M) \geq \left(1 - \frac{M-1}{N-1}\right).
\end{equation}
Thus combining (\ref{r5d}) and (\ref{r6d}), we get:
\begin{align}
\frac{R^D_s(M)}{R^*_s(M)} & \leq \frac{\frac{N-1}{M-1} - 1}{1 - \frac{M-1}{N-1}}\nonumber\\
													& = \frac{N-1}{M-1} \leq 17.
\end{align}
Thus we have proved that for any $N,K \in \mathbb{N}$ and all $\frac{N-1}{N} + 1\leq M\leq N$, there is a constant multiplicative gap of $17$ between the achievable secure decentralized rate and the information theoretic optimal for any secure scheme. It is to be noted that for $K>N$, the gap is unbounded in the regime 
\begin{align}
1< M < \frac{N-1}{N} + 1, 
\end{align} 
and scales with the number of users $K$. But $\frac{N-1}{N}<1$ for any $N$ and thus the regime of $M$ in which the gap is unbounded is in general negligible, especially when $N,K$ are large. This concludes the proof of Theorem \ref{th:6}. \qed

\bibliographystyle{IEEEtran}
\bibliography{arxiv_Final_version}

\end{document}